\documentclass{aa}
\usepackage[utf8]{inputenc}
\usepackage{graphicx}
\usepackage{txfonts}
\usepackage{url}
\usepackage[colorlinks=true,linkcolor=blue,citecolor=blue]{hyperref}
\usepackage{subfig}
\usepackage{xcolor}
\usepackage{amsmath}
\usepackage{amssymb}
\usepackage{booktabs}
\usepackage{mathtools}
\usepackage{caption}
\usepackage{floatrow}
\usepackage{float}
\usepackage{tikz}
\usepackage{tikz-3dplot} 
\usepackage{pgfpages}
\usepackage{listings}
\usepackage{lineno}
\usepackage{appendix}
\usepackage{natbib}
\bibpunct{(}{)}{;}{a}{}{,}

\newcommand\T{\rule{0pt}{2.6ex}}
\newcommand\B{\rule[-1.2ex]{0pt}{0pt}}
    
\begin{document} 

   \title{ExoDNN: Boosting exoplanet detection with artificial intelligence.}
   \subtitle{Application to Gaia Data Release 3}
   \author{
        A. Abreu\inst{1,2}
        \and J. Lillo-Box\inst{3}
        \and A. M. Perez-Garcia\inst{4}
        \and J. Sahlmann\inst{5}
        \and J. H. J. de Bruijne\inst{6}
        \and C. Cifuentes\inst{3}
        }

    \institute{
    ATG Science \& Engineering for the European Space Agency (ESA), ESAC, Spain \email{Asier.Abreu@ext.esa.int}
    \and
    Universidad Complutense de Madrid, Av. Complutense, s/n, Moncloa - Aravaca, 28040 (Madrid), Spain \email{asabre01@ucm.es}
    \and
    Centro de Astrobiolog\'ia (CAB, CSIC-INTA), ESAC campus, 28692, Villanueva de la Ca\~nada (Madrid), Spain
    \and 
    ISDEFE, Beatriz de Bobadilla, 3. 28040 Madrid, Spain  
    \and
    European Space Agency (ESA), European Space Astronomy Centre (ESAC), Camino Bajo del Castillo s/n, 28692 Villanueva de la Ca\~nada, Madrid, Spain
    \and
    European Space Agency (ESA), European Space Research and Technology Centre (ESTEC), Keplerlaan 1, 2201 AZ Noordwijk, The Netherlands
    }
    
   \date{Received June, 2025 / accepted October, 2025}
 
   \abstract
   {Transit and radial velocity (RV) techniques are the dominant methods for exoplanet detection, while astrometric exoplanet detections have been very limited thus far. Gaia has the potential to radically change this picture, enabling astrometric detections of substellar companions at scale that would allow us to complement the picture of exoplanet architectures given by transit and RV methods. }
   {Our primary objective in this study is to enhance the current statistics of substellar companions, particularly within regions of the orbital period–mass parameter space that remain poorly constrained by RV and transit detection methods.}
   {Using supervised learning, we trained a deep neural network (DNN) to recognise the characteristic distribution of the fit quality statistics corresponding to a Gaia Data Release 3 (DR3) astrometric solution for a non-single star. We created a deep learning model, ExoDNN, which predicts the probability of a DR3 source to host unresolved companions.}
   {Applying the predictive capability of ExoDNN to a volume-limited sample ($d\mathbf{<}100pc)$ of F, G, K, and M stars from Gaia DR3, we have produced a list of 7414 candidate stars hosting companions. The stellar properties of these candidates, such as their mass and metallicity,  are similar to those of the Gaia DR3 non-single-star sample. We also identified synergies with future observatories, such as PLATO, and we propose a follow-up strategy with the intention of investigating the most promising candidates among those samples.}
   {}
    \titlerunning{ExoDNN: boosting exoplanet detection with artificial intelligence.}
    \authorrunning{Abreu et al.}
    
    \maketitle
    \nolinenumbers
    
    \section{Introduction}\label{sect:intro}

    The current census of extrasolar planets contains more than 7000 objects\footnote{Statistics as of early 2025 from: \url{https://exoplanet.eu/home/}}, primarily compiled via transit and/or radial velocity (RV) detection methods. These techniques have helped expand our understanding of key physical properties of exoplanets, such as their radius and mass, enabling  studies of their bulk density and interior structure \citep[e.g.,][]{2007ApJ...669.1279S}. However, both the transit and RV methods suffer from known observational biases that favor the detection of close-in orbital companions \citep[e.g.,][]{Fischer_2014}.
    To reach longer period orbits, the astrometric detection method can be used \citep[e.g.,][]{Binnendijk1960,Quirrenbach_2011}. This involves detecting the reflex motion (wobble) of a star due to an orbiting companion (e.g., exoplanet). The size of this wobble, also known as astrometric orbit signature, scales inversely with distance; therefore, even nearby systems ($d\mathbf{<}100pc$) yield very small signatures (sub milli-arcsecond). This has resulted in just a handful of brown dwarf or exoplanet detections using astrometry to date \citep{2002ApJ...581L.115B,2005ApJ...630..528P,2013A&A...556A.133S,2022AJ....164...93C}. To this end, the ESA Gaia telescope \citep{2016A&A...595A...1G} has provided a unique opportunity for the astrometric detection and characterization of substellar companions to stars. Thanks to its unprecedented precision (end-of-nominal mission estimates for parallax and proper motion are 5$\mu$as and 3.5$\mu$as $yr^{-1}$, respectively, for $G$=3-12 mag \citealt{2014EAS....67...23D}), Gaia is expected to detect possibly thousands of new planetary and brown dwarf companions in long period orbits down to at least the giant planetary mass regime \citep{2008A&A...482..699C, 2014MNRAS.437..497S, 2014ApJ...797...14P} and up to a few hundred parsec over the whole sky. 

    In Gaia Data Release 3 \citep[][hereafter, DR3]{2023A&A...674A...1G}, the Gaia non-single-star pipelines \citep{2023A&A...674A...9H,2023A&A...674A..10H} have produced $\sim$170\,000 orbital solutions for binary systems, providing a first glimpse of the great potential of Gaia for stellar multiplicity analysis \citep{2023A&A...674A..34G}. A small subset of these solutions ($\sim$1800) correspond to sources with companion candidates in the substellar mass regime ($m\!<80 M_\text{Jup}$) (see Sections 5 and 8.1 in \citealt{2023A&A...674A..34G}). However, as these authors remark, Gaia DR3 processing limitations do not allow  new detections in the planetary mass regime ($m\!<20 M_\text{Jup}$) to be claimed solely on the basis of  astrometry. The next major Gaia data release, Gaia DR4, is expected to reach at least an order of magnitude higher astrometric precision than Gaia DR3, enabling the expansion of the sensitivity down to the planetary mass regime. Until then, Gaia DR3 can provide important insights onto the demographics of the substellar mass range covering $20 M_\text{Jup}\mathbf{\lesssim}m{\lesssim}80 M_\text{Jup}$ (see e.g., \citealt{2023MNRAS.526.5155S,2023A&A...680A..16U}). The rich dataset available in DR3, is also well suited for machine learning applications \citep{2024A&A...682A...9G,2023A&A...675A..68V,2025A&A...693A.268R} and, since the astrometric time series measurements used to produce the DR3 astrometric solutions are not yet publicly available, machine learning can complement the classical orbital fitting method for identification of substellar companions \citep[e.g.,][]{2025MNRAS.537.1130S}.
     
    In this work, we present a deep learning framework for identifying companion candidates in Gaia DR3 astrometry. Although deep learning has been extensively applied to transit detection \citep[][]{2018AJ....155...94S,2018ApJ...869L...7A,2019AAS...23314016D,2022ApJ...926..120V,2022AJ....163..237D}, its use in astrometric companion detection remains largely unexplored. Our method is designed to (i) select those DR3 parameters that effectively trace binarity; (ii) model their complex non-linear relationships using a deep neural network, while mitigating any over-fitting; and (iii) ensure model interpretability, so that predictions can be linked to astrophysical indicators. The resulting model, called ExoDNN, identifies non-single star candidates in Gaia DR3 with high reliability using astrometric fit-quality statistics and other DR3 parameters available in the main source table \citep{Hambly2022}.
    
    The paper is structured as follows. Section 2 introduces the ExoDNN algorithm and describes the data and methods used for its training and validation. In Section 3, we describe how we applied ExoDNN to a subset of Gaia DR3 stars and present a preliminary list of candidate stars to host companions. Section 4 discusses the results, while Section 5 summarizes the conclusions, highlighting synergies with upcoming exoplanet missions and outlining prospects for follow-up observations.
     
    \section{The ExoDNN algorithm}
    \subsection{Detection probability and statistical indicators}\label{sect:agis_statistics}

    The astrometric signature of a binary system is typically defined as an angular measure  $\alpha$(arcsecond)=$a_0/d$, where $a_0$ is the semimajor axis of the orbit described by the star around the barycentre of the system (in AU) and $d$ is the distance to the observer (in pc). To determine the probability $p$ of detecting this signature, we can establish a detection criterion based on achieving a specific signal-to-noise ratio (S/N=$\alpha/\sigma_m)$, where $\sigma_m$ is the measurement uncertainty (ignoring other possible sources of noise). The S/N will then effectively be a cut-off in probability space, that is, serve as a criterion for acceptable confidence level in the detection (and, thus, $p\propto S/N$). For the case of unknown orbital periods, \cite{2008A&A...482..699C} used this criterion and derived a threshold of $S/N>3$, for a successful astrometric orbit detection at 95\% confidence level. Using a similar criterion, \cite{2015MNRAS.447..287S} took into account the number of measurements ($N_m$) to show that a $\sqrt{N_m} \times S/N \ge 20$ is necessary for detection when using astrometric data only.

    However, the astrometric signature ($\alpha$) is not directly observable by Gaia. Instead, Gaia measures the one-dimensional (1D) position (abscissa) of a star in the along-scan direction, which we hereafter refer to  as $\eta$. Gaia data reduction pipeline, the Astrometric Global Iterative Solution \citep[AGIS, hereafter]{2012A&A...538A..78L,2021A&A...649A...2L}, uses a single star model to compute the expected along-scan position of a source in the sky at time $t$ ($\eta^{\text{exp}}$) and then minimizes the astrometric residuals (observed minus expected) ($\eta^{\text{obs}}-\eta^{\text{exp}}$) to generate an astrometric solution. 
    \begin{table}[h]
    \caption{AGIS astrometric fit quality statistics}
    \label{tab:agis_statistics}
    \small
    \centering
    \begin{tabular}{c c}
    \textbf{AGIS astrometric fit statistic} & \textbf{Description}\\
    \hline
    ruwe & re-normalized unit weight error \T \\
    astrometric\_gof\_al (${gof}_{AL}$)  & goodness-of-fit statistic \\
    astrometric\_excess\_noise ($\epsilon$) & excess noise of the source \\
    astrometric\_chi2\_al ($\chi^2$)  & astrometric goodness-of-fit \\
    astrometric\_sigma5d\_max ($\sigma_{5d}$) & longest semimajor axis of the \\ 
     & 5D error ellipsoid \B \\
    \hline
    \end{tabular}
    \tablefoot{A detailed description of these statistics is available in \cite{2021A&A...649A...2L} Section 5.1.}
    \end{table}
    This astrometric solution provides the estimated (best-fit) astrometric parameters of the source, along with a set of fit quality statistics (see Table~\ref{tab:agis_statistics}). In a binary system, the excess motion induced by the unseen companion will affect the observed position of the star $\eta^{\text{obs}}$, causing the astrometric residuals to be larger than in the single star case and  resulting in increased values of fit $\chi^2$. This has been used by several authors in the past to identify solutions that deviate from the single star model used by AGIS. As an example \cite{2020MNRAS.496.1922B} and  \cite{2022MNRAS.513.5270P} used the re-normalized unit weight error (\texttt{ruwe}), which is proportional to the $\chi^2$ of the astrometric fit, to search for astrometric binaries in the Gaia catalogue.  \cite{2021ApJ...907L..33S} found a strong correlation between this statistic and the photocentric orbit size of eclipsing binaries. The astrometric excess noise ($\epsilon$) has also been used by  \cite{2022MNRAS.510.3885G} in search for X-ray binaries. 
    
    \noindent However, it should be noted that variability \citep{2020MNRAS.496.1922B}, stellar crowding \citep{2023A&A...677A.185L}, or the presence of a proto-planetary stellar disk \citep{2022RNAAS...6...18F} has been shown to contribute to high values of $\texttt{ruwe}$ without a binary cause.
    Besides, both statistics are known to be affected by calibration errors that make them sensitive to the magnitude, color of the source (etc.); although these effects have been mitigated by construction in $\texttt{ruwe}$, they still affect $\epsilon$ in DR3 \citep[][Section 5.3]{2021A&A...649A...2L}. Therefore, these caveats must be taken into account when using AGIS statistics as probes for the presence of unseen companions.

    \subsection{Deep learning}\label{sect:deep_learning}
    
    Despite the limitations mentioned earlier, it is clear that binarity has an effect on the fit quality statistics provided by AGIS, so these can be used to predict the probability, $p$, that a source from Gaia DR3 is non-single, given an array ($\bar{x}$) of values for those statistics. Now, since we lack an explicit form of the relation between the astrometric signature and the fit quality statistics, finding an analytical form of the probability $p(\bar{x})$, can be challenging. Therefore, we decided to take a machine learning approach and use a special type of artificial neural network, a  deep neural network (DNN), to model $p(\bar{x})$. 
    
    In general, artificial neural networks can approximate any continuous function to an arbitrary degree of accuracy \citep{HORNIK1989359}, and learn such an approximation from data, using provided examples in a setting called supervised learning. If we use a dataset, $\mathcal{D}=\{(\bar{x}_1,y_1),\dots,(\bar{x}_k,y_k)\},$ composed by pairs of examples ($\bar{x}_i,y_i$), where $\bar{x_i}$ is array of AGIS fit statistics with known values for a source and $y_i$ only has two possible values ($y_i$=0 representing a single star, and $y_i$=1 representing a system with a companion detectable by Gaia), we can teach the network the underlying mapping between $\bar{x}$ and $y$. The probability density of the random variable $y$ can be described by a continuous Bernoulli distribution \citep{loaiza2019continuous}, expressed as 
    \begin{equation}
        p(y\vert x;\theta) = \theta^x (1 - \theta)^{1-x}
    ,\end{equation}
    \noindent where $\theta \in$ [0,1] is an unknown parameter that maximizes the probability that $y\mathbf{=}y_i$ given $\bar{x_i}$, and $\bar{x_i}\mathbf{\in}[0,1]^N$. The DNN defined in this manner will effectively learn the parameter of this distribution and produce the output we are looking for, namely, an estimate of the probability $p(\bar{x})$ over many Bernoulli trials, each with a probability expressed as $\theta$.
    
    \subsection{Generation of the training set}\label{sect:simulation}
    
    To train ExoDNN, we need example astrometric solutions for the positive case (stars with a detectable companion) and for the negative case (stars with no detectable companion). Given that a homogeneous, sizeable sample of bona fide single stars was not available for our study, we resorted to compiling a set of synthetic data.
    
\begin{figure}[h!]
    \includegraphics[width=0.9\textwidth]{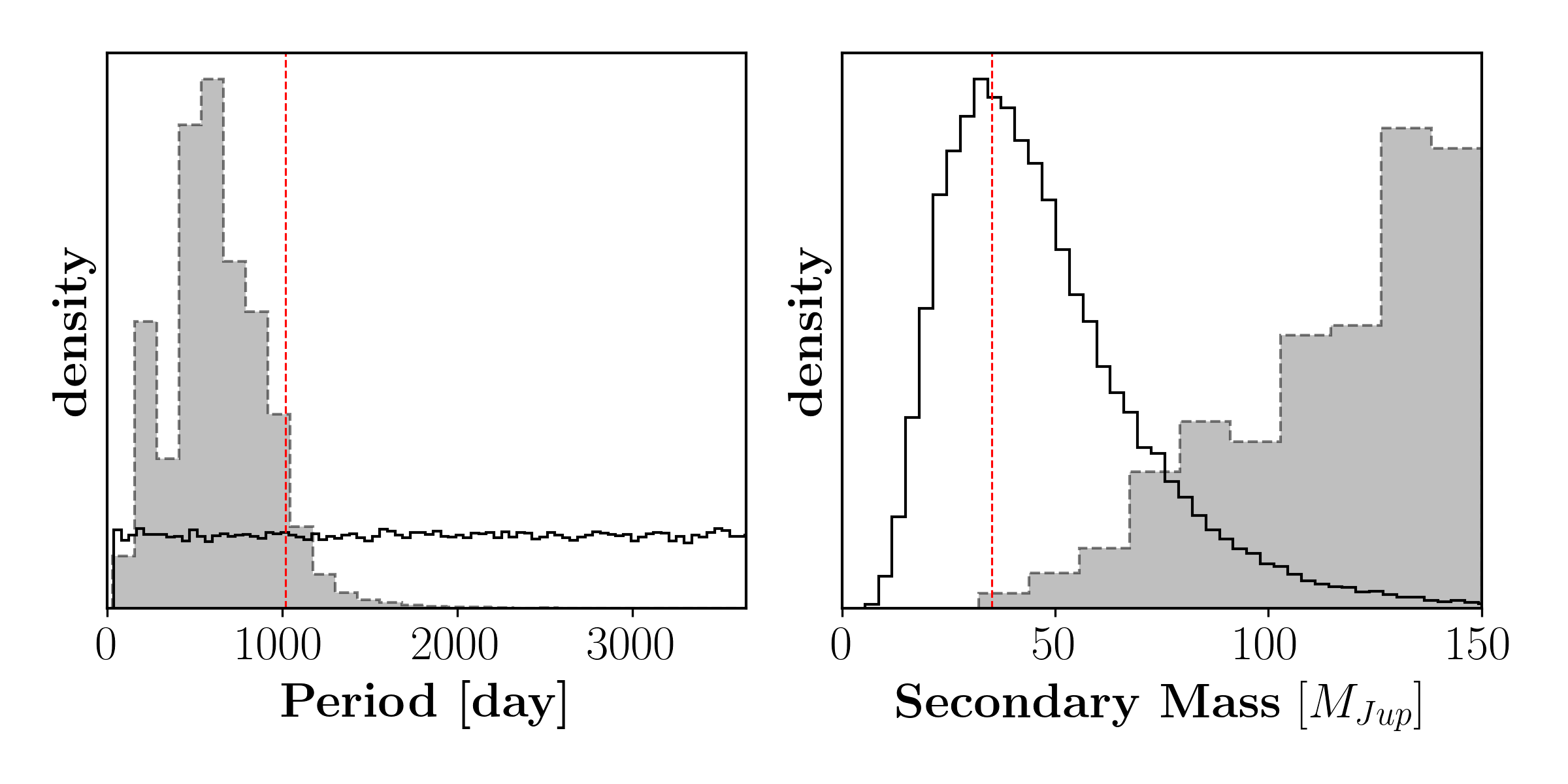}%
    \caption{LEFT: Simulated orbital periods (black) compared to the orbital periods from DR3 NSS Orbital solutions (shade), with the DR3 time baseline marked in vertical. RIGHT: Simulated secondary masses (black), compared to the catalogue of DR3 binary masses (shade), with the peak of the distribution of simulated secondary masses in vertical.}
    \label{fig:simulation_distros}
    \end{figure}

    We generated 100\,000 examples with roughly equivalent proportion ($\sim$50\%) of single stars and detectable binary systems, to avoid biasing the neural network towards preferential detection of stars from an over-represented group. These systems were placed randomly in the sky at uniform distances in the range ($1\!\le\!d\!\le\!100$pc). Primary and secondary masses were extracted from the log-normally distributed ranges ($0.5 M_\odot\!\le\!M_1\!\le\!1.5M_\odot$) and ($10 M_\text{Jup}\!\le\!M_2\!\le\!150M_\text{Jup}$), respectively; the latter roughly covering a mass range from giant planets to low-mass stars \citep{2001MNRAS.322..231K,2011ApJ...727...57S,2014prpl.conf..619C}. We note that our secondary masses (see Figure~\ref{fig:simulation_distros}) are  generally lower than the $\sim$1800 candidates from the DR3 NSS catalogue of binary masses \citep[Section 5]{2023A&A...674A..34G}. This is intentional, as our goal is to simulate astrometric signatures induced by very low-mass companions. 
    \begin{figure}[h!]
    \caption{TOP: Barycentre motion of a simulated single star corresponding to Eq.~\ref{eqn:single_star} with best fit astrometric parameters. The observed position of the source is marked with solid white circles, and the 1D along-scan observations with gray circles.  BOTTOM: Same, but for a simulated binary system, where the observed position of the source is perturbed by a companion (exaggerated for illustration purposes).}
    \includegraphics[width=0.9\textwidth]{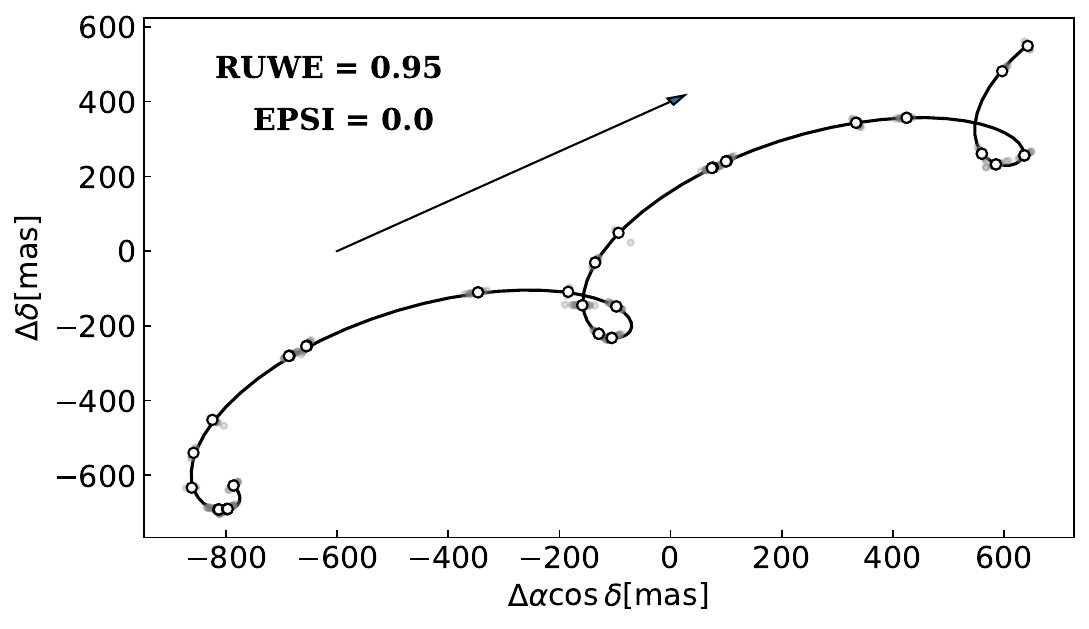}
    \vspace{10pt}
    \includegraphics[width=0.9\textwidth]{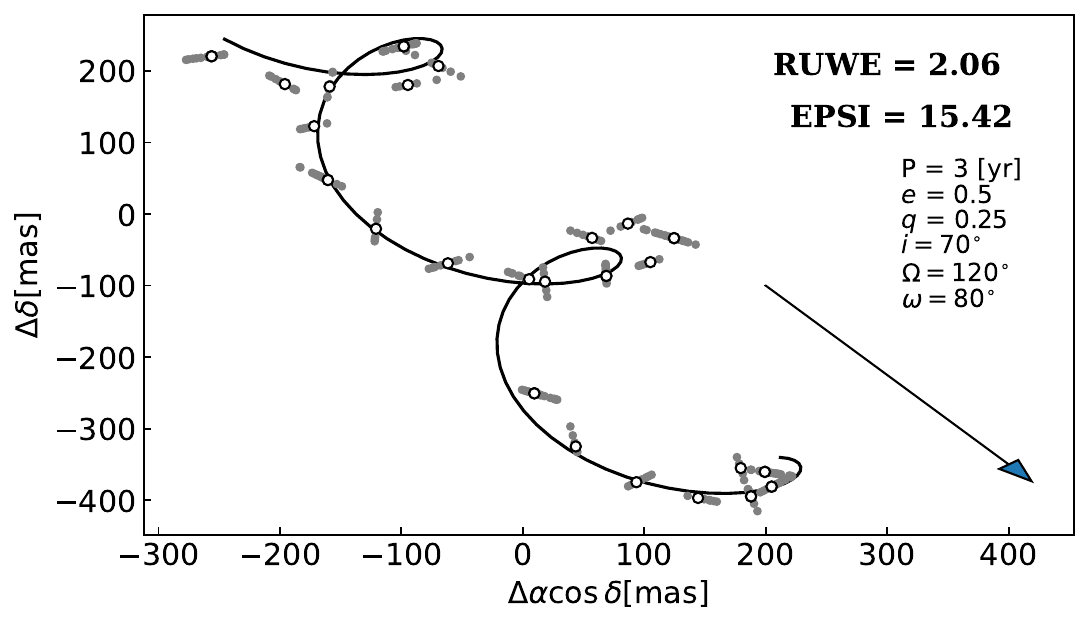}
    \label{fig:orbits}
    \end{figure}
    The orbital periods were drawn from the uniform range ($0.1\text{yr}\!\le\!P\!\le\!10\text{yr}$), which is the maximum baseline of Gaia observations for the full mission. Once total mass and period of the system were defined, these determined the orbit size ($a$) through Kepler's third law. The remaining orbital parameters for the Keplerian companion were the eccentricity ($e$), orbital inclination ($i$), argument of the periastron ($\omega$), longitude of the ascending node ($\Omega$), and time at the periastron  ($t_p$). These parameters were randomly drawn from the uniform distributions: $0\!\le\!e\le\!1$, $0\!\le\!i\!\le\!90^{\circ}$, $0\!\le\!\omega\!\le\!360^{\circ}$, $0\!\le\!\Omega\!\le\!180^{\circ}$, and $0\!\le\!t_p\!\le\!P$. 

    We then used the \texttt{Astromet} library from \cite{2022MNRAS.513.5270P} to generate astrometric time series for both single and binary simulated system. Similarly to AGIS, \texttt{Astromet} uses a five-parameter (5$p$) astrometric model to describe the motion of a single source. This model depends on right ascension ($\alpha^*\!=\!\alpha \cos\delta$), declination ($\delta$), the proper motions ($\mu_{\alpha^*}$, $\mu_\delta$), and the parallax ($\varpi$) of the star,  expressed as 
    \begin{align}
    \label{eqn:single_star}
     \eta(t) = [\alpha^* - \alpha_0^* + \mu_{\alpha^*}\;(t-t_0) ] \sin \phi_t\\ 
     +\;[\delta - \delta_0 + \mu_{\delta}\;(t-t_0)] \cos \phi_t \notag \\
    + \varpi \;f_{\varpi}, \notag
    \end{align}
    
    \noindent where ($\alpha_0^*$, $\delta_0$) correspond to the on-sky position at time $t_0$, $\phi_t$ is the position angle of the scan at time $t$, and $f_{\varpi}$ is the source along-scan parallax factor ($\partial \eta / \partial \varpi$). $\phi_t$ and $f_{\varpi}$, were obtained from the Gaia scanning law\footnote{Gaia scanning law for the full mission can be obtained at: \url{https://www.cosmos.esa.int/web/gaia/full-mission-scanning-law}}. 
    
    With the along-scan positions defined, we were able to simulate the 1D observations ($\eta^\text{obs}_{j}$) that Gaia would perform on source $j$ over a time range equivalent to the Gaia DR3 time baseline of 34 months. Each simulated observation can be expressed as $\eta^\text{obs}_{j}(t)\mathbf{=}\eta_{j}(t)+\sigma_m$, where $\eta_{j}(t)$ is given by Eq.~\ref{eqn:single_star}, and $\sigma_m$ is the along-scan single measurement uncertainty. The real Gaia observation uncertainty displays certain dependencies on time, the across-scan rate, and magnitude (see \citealt{2021A&A...649A...2L} Appendix A); however, for simplicity, we chose a Gaussian measurement error with zero mean and 50$\mu$as standard deviation, the latter corresponding to the median formal DR3 measurement uncertainty at G=12 (again \citealt{2021A&A...649A...2L} Appendix A). We then fit the single star model from  Eq.~\ref{eqn:single_star} to our simulated observations and generated a five-parameter astrometric solution for each star in our simulation (see example in Figure~\ref{fig:orbits}). The least-squares fit implemented by \text{Astromet} (see \citealt{2020MNRAS.495..321P} Appendix D), closely follows AGIS implementation, resulting in astrometric solutions that contain the exact same set of fit quality statistics as provided by AGIS. The choice of fitter is therefore very convenient, because a model trained on those astrometric solutions will be applicable to the five-parameter astrometric solutions available in DR3 main source table.  
   
    At this stage, however, we only had (5$p$) best fit astrometric solutions ($\hat{\alpha}$, $\hat{\delta}$, $\hat{\mu}_{\alpha}$, $\hat{\mu}_{\delta}$, $\hat{\varpi}$), along with their uncertainties and the corresponding fit quality statistics for each solution. To avoid biasing the model towards any positional, motion or distance preference, we dropped the five astrometric solution parameters, and only kept their uncertainties and the fit quality statistics. Also, to have more representative data of the actual DR3 parameter space, we decided to add 16 photometric parameters and 4 spectroscopic DR3 parameters, by conditionally sampling distributions of those parameters derived from real DR3 sources. Details and caveats on the derivation of photometric and spectroscopic information are provided in Appendix~\ref{appendix:additional_data} and omitted here for brevity. 
    
    This process resulted in a labelled dataset composed of 100\,000 examples ($\bar{x}_i,y_i$), where $\bar{x}_i$ was an array of 31 DR3 parameters, and $y_i$ was a value (label) denoting either a binary system with a Gaia detectable companion ($y_i\mathbf{=}1$) or a single star ($y_i\mathbf{=}0$). The final list of parameters used to train our model is provided in Table~\ref{tab:model_params}.

    \subsection{Preprocessing}

    We then scaled robustly the training dataset using the Scikit-Learn package \citep{JMLRv12pedregosa11a}. To do this, we centred each model parameter by subtracting its median and then dividing by its interquartile range. Robust scaling is a commonly used scaling technique \citep[e.g.,][]{geron2017handson}  which takes into account the possible presence of outliers and avoids that the different scales naturally present in the data bias the model towards a particular parameter. The training dataset was then split using a 70-15-15\% scheme. 70\% of the whole data containing positive and negative examples were used to train our network (training set), 15\% (validation set) were used to tune the network hyper-parameters and the remaining 15\% (the test or hold-out set) were left aside to check how well the model generalizes, this is, how well it predicts over unseen data. Shuffling of the whole data was also performed before the train-val-split to ensure adequate mixture of positive and negative examples across the train, test and validation data splits.

    \begin{table}[!ht]
    \small
    \caption{Gaia DR3 parameters used to train the model.}
    \label{tab:model_params}
    \centering
    \begin{tabular}{ll}
    \textbf{Parameter type} & \textbf{Parameter name} \\
    \hline \\
    astrometry  & astrometric\_n\_obs\_al\\
                & ra\_error  \\
                & dec\_error \\
                & parallax\_error\\
                & pmra\_error\\
                & pmdec\_error\\
    fit statistic& ruwe \\
                & astrometric\_excess\_noise\\
                & astrometric\_chi2\_al\\
                & astrometric\_gof\_al\\
                & astrometric\_5d\_sigma\\
    photometry  & phot\_g\_mean\_flux \\
                & phot\_g\_mean\_flux\_error\\ 
                &phot\_g\_mean\_flux\_over\_error\\
                &phot\_g\_mean\_mag\\
                &phot\_bp\_mean\_flux\\
                &phot\_bp\_mean\_flux\_error\\ 
                &phot\_bp\_mean\_flux\_over\_error\\ 
                &phot\_bp\_mean\_mag\\
                &phot\_rp\_mean\_flux\\
                &phot\_rp\_mean\_flux\_error\\
                &phot\_rp\_mean\_flux\_over\_error\\
                &phot\_rp\_mean\_mag\\
                &phot\_bp\_rp\_excess\_factor\\ 
                &bp\_rp \\
                &bp\_g\\
                &g\_rp\\
    spectroscopy&radial\_velocity\\
                &radial\_velocity\_error\\ 
                &grvs\_mag\\
                &grvs\_mag\_error\B\\
    \hline
    \end{tabular}
    \tablefoot{A Detailed description of each of them can be found at: \url{https://gea.esac.esa.int/archive/documentation/GDR3/Gaia_archive/chap_datamodel/sec_dm_main_source_catalogue/ssec_dm_gaia_source.html}.}
    \end{table}

    \subsection{Model training}\label{sect:model_training}

    \subsubsection{Model architecture}

    ExoDNN is composed of an input layer followed by three densely connected hidden layers, one dropout layer and the final output layer. The input layer contains 31 neurons to map our 31 DR3 parameters. The hidden layers have 64 neurons each. The neurons on these layers use a rectified linear unit (ReLU) gate, as it speeds the training in relative comparison to other non-linear gates. The role of the dropout layer is to reduce over-fitting \citep{JMLR:v15:srivastava14a}, a situation in which a model produces a perfect prediction over the data it was trained against, but fails to generate good predictions when exposed to unseen new data. Finally, to ensure that a valid probability is produced by the network, the output layer contains a single neuron with a sigmoid activation function which always produces a predicted probability value $\hat{p}$ in the range $[0,1]$. 
    \begin{table}[h!]
    \small
    \caption{Neural network hyper-parameters.}
    \label{tab:network_architecture}
    \centering
    \begin{tabular}{l l l}
    & \textbf{Hyper-parameter} & \textbf{Value} \\
    \hline
    Architecture    & No. of hidden layers       & 3 \T\\
                    & Neurons per hidd. layer    & 64 \\
                    & Hidd. Neuron gate          & ReLU \\
                    & Hidd. layer type           & Dense \\
                    & Output layer gate          & Sigmoid \\ 
    Training        & Data split                 & 70-15-15\% \\
                    & seed                       & 42 \\ 
                    & shuffle                    & true \\                 
                    & Learning rate              & 10$^{-4}$ \\ 
                    & Batch size                 & 64 \\
                    & Optimizer                  & Adam \\
                    & Loss function              & Negative log-loss \\
    Overfitting     & Dropout rate               & 0.2 \\ 
                    & EarlyStopping(monitor)     & val data loss \\
                    & EarlyStopping(patience)    & 10 \B\\
    \hline
    \end{tabular}
    \end{table}
    \noindent We explored network architectures with two to five hidden layers. Configurations with two layers did not achieve the required precision, whereas five layers increased training time without measurable performance gains. The number of neurons per layer was varied between 32 and 256, guided by standard heuristics that balance complexity and computational efficiency. A configuration with 64 neurons per layer yielded the most stable convergence and was therefore adopted. A detailed description of many of the concepts presented in this section such as architecture design, network training, activation gates, back propagation, over-fitting and performance metrics, can be found in \cite{Goodfellow2016}.
    
    \noindent Table~\ref{tab:network_architecture} summarizes the final architecture of ExoDNN, along with the training hyper-parameters (optimizer, learning rate, mini-batch size, dropout rate and stopping criterion). To implement ExoDNN we used the Tensorflow package \citep{tensorflow2015large}.
    \subsubsection{Model optimization}\label{sect:optimization}
    
    A neural network learns by iteratively adjusting its internal parameters (weights and biases) through forward and back propagation of the training data. During forward propagation, the network makes predictions (using small batches of data) of the probability $\hat{p}$ that a training example $\bar{x_i}$ in the batch belongs to the positive class ($y_i\mathbf{=}1$). Then, these predictions are compared to the true class using a loss function that quantifies how far off the predictions are from the actual labels $y_i$. Our loss function is the average negative log-likelihood of the training set,
expressed as    \begin{equation}\label{eq:log-loss}
     l(\theta) = -\frac{1}{N} \sum_{i} {y_i\times \log(\theta) + (1 - y_i)\times \log(1-\theta)}
    \end{equation}
    
     \noindent where $N$ is the number of training samples and $y_i$ is the actual class of the $i$-th sample. The variable $\theta$ is the probability computed by the network that the training sample $i$ belongs to the positive class. We note that minimizing the negative log-likelihood is equivalent to maximizing the likelihood of the training data under our parametrized model of the random variable $y$. During back-propagation, the weights are re-adjusted to lower the loss on the training set. The algorithm responsible for re-adjusting the weights is called optimizer and we chose the Adam optimizer \citep{2015-kingma} due to it's computational efficiency and fast convergence. The model performs multiple forward and backward passes on iterations called epochs. During an epoch a full pass through the entire training set is performed, and this process is repeated until a stopping criterion is met. In our case, we used a stopping criterion called "early stopping" that halts the training when it detects no improvement on the loss computed over the validation set after a successive number of epochs. This choice was also aimed at avoiding a network over-fitting.  
     
\begin{figure}[!h]
    \centering
     \includegraphics[width=9cm]{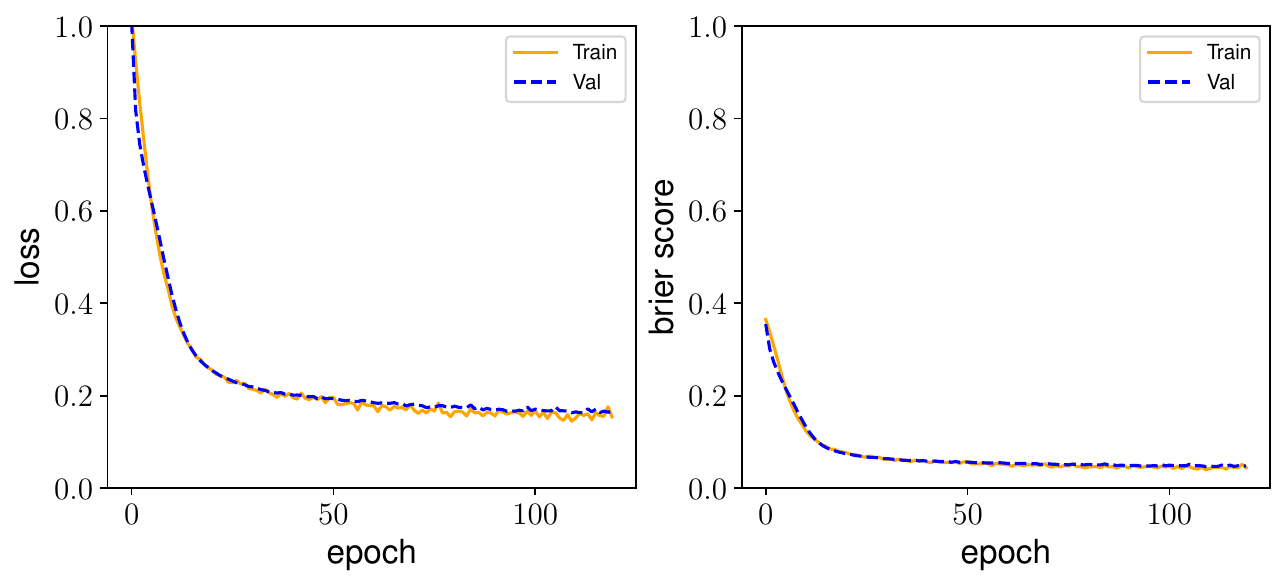}
    \caption{LEFT: Evolution of the training and validation set losses over the different epochs. RIGHT: Evolution of the Brier-score during the training.}
    \label{fig:model_training}
    \end{figure}

    The training process explained earlier is designed to teach ExoDNN to produce probability estimates that are as close as possible to the true underlying class probabilities. Therefore, to evaluate the model performance, we selected a metric called the Brier-score, which is equivalent to the mean-squared error but for a probabilistic classification. This metric has values higher or equal than zero, and is defined as
    
    \begin{equation}
    Brier = \frac{1}{N} \sum_{i=1}^{N}({\hat{p}_i-y_{i}})^{2}
    \label{eq:metrics}
    ,\end{equation} 

    \noindent where $N$ is the number of examples, $\hat{p}_i$ is the probability computed by our model for example, $\bar{x_i}$, and $y_i$ is the actual value of the random variable $y$ for that example. Figure~\ref{fig:model_training} shows the evolution of the log-loss (left) and brier-score (right) during the training. 
    Since ExoDNN outputs probabilities, to evaluate its performance over a test set with "hard" labels (positive or negative class label), we first needed to establish a classification threshold. This threshold determines when a given predicted probability $\hat{p}_i$ belongs to the positive class (binary star). The optimal threshold was computed by making predictions over our hold-out (test) set while varying the probability threshold and finding which of those thresholds maximized a chosen binary classification metric. The selected metric was the $f1$-score metric because it represents a trade-off between precision and recall. In this context it is important that we correctly classify stars that host detectable companions as such (precision), and that we capture as many of those as possible (recall). The $f1$-score has values between zero and one, and is defined as 
    \begin{equation}
    f1=\frac{2*\text{Precision}*\text{Recall}}{\text{Precision}+\text{Recall}}
    .\end{equation}

    \begin{figure}[!h]
    \centering
    \includegraphics[width=9.0cm]{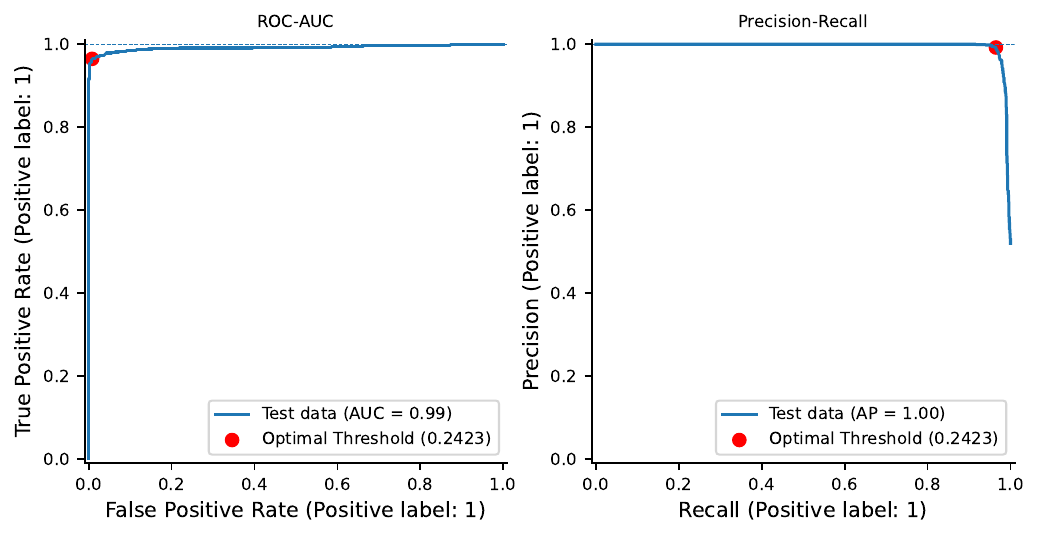}
    \caption{LEFT: classification threshold computation using the area under curve. RIGHT: classification threshold computation using the precision-recall curve. The optimal threshold that achieves the best compromise between false positive and true positive rates is marked as a red dot. This is our chosen $p_0$=0.242.}
    \label{fig:precission_recall}
    \end{figure}
    
    \noindent ExoDNN retained a very low false positive rate as the classification threshold was varied, as shown in Figure~\ref{fig:precission_recall} (left), and also maintained high precision and recall at the same time, as shown in Figure~\ref{fig:precission_recall} (right). The optimal balance between the true positive and the false positive rates over the test data was found for a probability threshold value of $p_0$=0.242. This means that a given example is considered to belong to the positive class, this is, star possibly hosts Gaia detectable companion(s), when the predicted probability on output, $\hat{p}$, is strictly higher than $p_0$, $\hat{p}(\bar{x_i})\mathbf{>}p_0$. 

    \subsubsection{Model evaluation}
    
    Using the 15\% hold-out set from the train–validation–test split, we evaluated the model on unseen data. According to our performance metrics (Section \ref{sect:optimization}), a well-performing model should exhibit a low Brier score (approaching zero) and a high $f1$-score (approaching one). For benchmarking, we compared our model against commonly used binary classifiers: the Logistic Regression \citep{cox1958regression} and the Random Forest \citep{2001MachL..45....5B}, a couple of classifiers that are commonly used for binary classification. A good overview of these models is available in~\citep[Chapters 4,11,15]{hastie01statisticallearning_neural_nets} and details on their parameter settings are provided in Appendix~\ref{appendix:additional_models}. Table~\ref{tab:classification_metrics} shows a summary of the performance metrics obtained by ExoDNN and the alternative classifiers. We found that ExoDNN outperformed the alternative binary classifiers, showing both a low Brier-score and a high $f1$-score, supporting its adoption over standard classifiers.
    
    \begin{table}[h!]
    \caption{Model evaluation results.}
    \label{tab:classification_metrics}
    \centering
    \begin{tabular}{c c c}
    \textbf{Model} & \textbf{$f1$-score} & \textbf{Brier score}\\
    \hline
    ExoDNN               & 0.901 & 0.034 \T\\
    Logistic Regression  & 0.817 & 0.125 \\
    Random Forest        & 0.812 & 0.159 \B\\
    \hline
    \end{tabular}
    \tablefoot{Binary classification scores for ExoDNN and alternative classifiers over the hold-out dataset.}
    \end{table}

    \subsection{Model inference and interpretation}\label{sect:model_inference}
    
    We  describe above  how ExoDNN was built and trained, but we might also consider how  it comes up with a predicted probability value for a given input. Our model has learned a non-linear approximation $f$ (using weights and biases) of the true conditional probability, $P(y\mathbf{=}1\vert \bar{x}_i)$, where $\bar{x}_i$ is an array of 31 DR3 parameters for a given star. When asked to predict over the input $\bar{x}_i$, it will produce an estimate $\hat{p}$ by computing $\hat{p}(\bar{x}_i)=\sigma(f(\bar{x}_i))$, where $\sigma$ is the sigmoid function,
    \begin{equation}
        \hat{p}(\bar{x}_i)=\frac{1}{1+e^{-f(\bar{x}_i)}}
    .\end{equation}
    \begin{figure}[h!]
    \includegraphics[width=9cm]{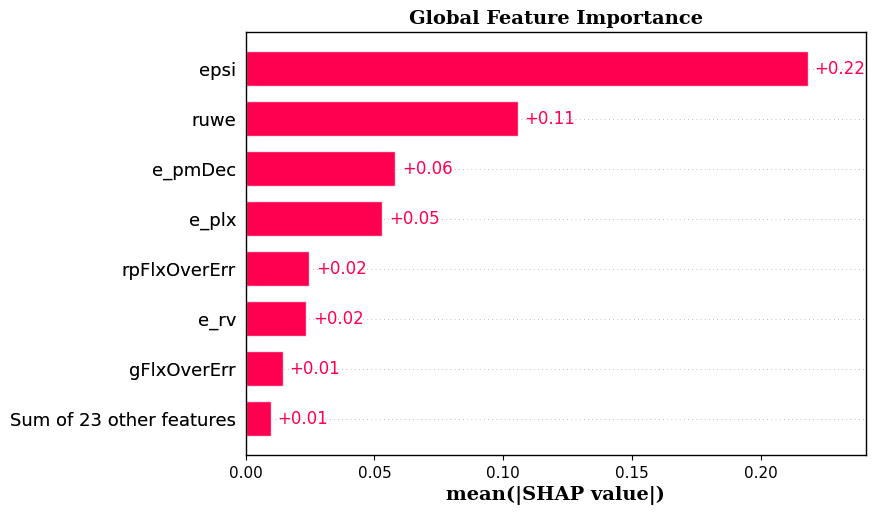}%
    \vspace{10pt}
    \includegraphics[width=9cm]{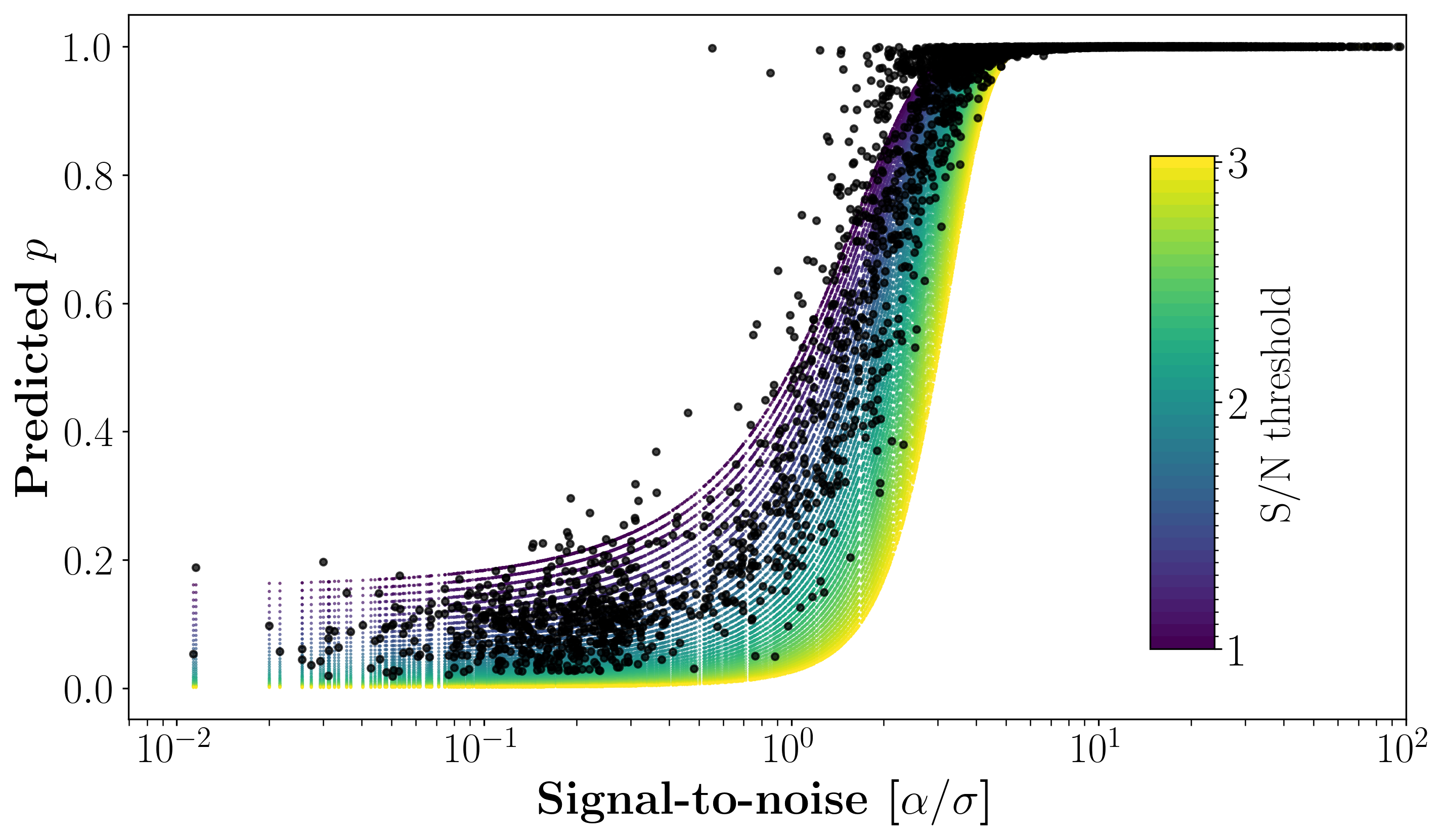}%
    \caption{TOP: Parameter impact on the model output as estimated by the neural network model ranked by decreasing impact. High parameter values are color-coded in red, while low values are displayed in blue. BOTTOM: Predicted probability for each binary star example on the test dataset (black dots), together with a detection probability proxy computed as $p=1-\Phi(S/N,\delta)$), where $\Phi$ is the CDF of the normal distribution, $S/N=\alpha/\sigma$, and $\delta$ is a variable S/N threshold from 1 to 3.}
    \label{fig:model_parameters}
    \end{figure}  
    
    \noindent Now, while the model will use all 31 DR3 parameters to compute $\hat{p}$, not all of them have the same weight towards that computation. To better understand the relative importance of those 31 DR3 parameters in the prediction, we used SHapley Additive exPlanations (SHAP) \citep{lundberg2017unified}, a commonly used tool to interpret machine learning models.  Figure~\ref{fig:model_parameters} (top) shows the relative importance of the most influential parameters towards the model prediction. The x-axis corresponds to the mean shap value. This is a quantitative measure of how much each feature changes the prediction compared to the average prediction, and gives an idea of its relative importance against other features.  We can see how $\texttt{ruwe}$ and $\texttt{astrometric\_excess\_noise}$ ($epsi$) have the highest relative importance for ExoDNN prediction, which is expected based on our analysis on how binarity affects the astrometric fit statistics (Section~\ref{sect:agis_statistics}). Then we have a group of parameters that trace the formal uncertainties in the astrometric solution, the proper motion (declination) and parallax errors ($pmDecErr,plxErr$), which are also expected to increase due to deviation from a purely linear motion model in the binary case. The RV error ($radVelErr$) traces the periodic velocity wobble induced by an orbiting companion in the host star, and therefore we would also expect this parameter to have a significant impact on the model decision. Finally, we have a group of photometric uncertainty related parameters ($rpFlxOverErr$ , $gFlxOverErr$).  \cite{2025AJ....169...29S} show how unresolved binarity has a direct impact on flux contamination over the Gaia detection window, resulting in larger photometric residuals. The remaining 23 model parameters were still used, but we note that they have less of an impact in the computation of $\hat{p}$. 
    
    To complement this analysis, we display in Figure~\ref{fig:model_parameters} (bottom) the computed probability $\hat{p}$ for each of the simulated binaries in the test data sample, along a detection probability proxy computed for different S/N levels. We can see how $\hat{p}$ scales with the astrometric S/N. This is also reassuring with respect to ExoDNN's behavior, indicating that the model has learned a proxy for the detection probability based on the astrometric S/N ratio noted in Section~\ref{sect:agis_statistics}.  

    \subsection{Model testing}\label{sect:model_test}
    
    The hold-out dataset used to check ExoDNN generalization (see Section~\ref{sect:model_training}), is fully synthetic. However, such data cannot capture the intrinsic DR3 calibration features mentioned in Section~\ref{sect:agis_statistics}. Therefore, we used unresolved binary systems detected by Gaia in DR3 (true positives) to test the behavior of ExoDNN when exposed to unseen and real data. These are: (1) the 384 DR3 NSS OrbitalTargetedSearch* and OrbitalAlternative* solutions (NSS ORBT) from \citep[Table 1]{2023A&A...674A..34G}, within 100pc; (2) the 1332 DR3 NSS AstroSpectroscopic solutions (combined single lined spectroscopic and astrometric binary) (NSS ASB1) from \citealt[Table 1]{2023A&A...674A..34G}, within 100pc; (3) the 48 M-dwarfs from the CARMENES input catalogue \citep{Cifuentes2025A&A...693A.228C} that have have a corresponding DR3 NSS solution. For all these systems, we had access to an existing orbital solution either from literature or from the Gaia DR3 NSS solutions, with an estimate of the size of the semimajor axis of the photocentric orbit ($a_0$). We then adopted the astrometric S/N detection criterion from \cite{2015MNRAS.447..287S}, namely: $\sqrt{N_{\text{obs}}}\,\alpha/\sigma_{\eta}\mathbf{>}20$, where $\alpha\mathbf{=}a_0/d$, $N_{\text{obs}}$ corresponds to the \texttt{astrometric\_n\_good\_observations\_al} DR3 parameter of the given source, and $\sigma_\eta\mathbf{=}150~\mu$as is the median along-scan measurement uncertainty estimated for DR3 sources at $G\mathbf{=}12$ (see Figure~3 in \citealt{2023A&A...674A..10H}). 
    
    This computation is merely a consistency check, since by construction all the selected real sources induce enough astrometric signal to be detectable by Gaia. Therefore, the number of expected detections on the given test sample is the same as the total number of sources for that sample. Finally, to measure ExoDNN baseline false positive rate, we generated a further test sample composed of 5000 new synthetic single stars using the approach described in Section~\ref{sect:simulation}. A perfect model should predict close to zero probability of hosting a companion for these true negative sources. The false positive rate, as determined from the control sample, resulted in approximately 1.2\%. 
    
    Table~\ref{tab:external_testing} shows the summary results of this process over the entire set of samples. Our model accuracy (detected/expected) varied depending on the sample, from $\sim$70\% for the CARMENES M-dwarf sample, to over 90\% for the Gaia NSS single lined astro-spectroscopic binaries. 
    \begin{table}[h!]
    \caption{Model testing results.}
    \label{tab:external_testing}
    \centering
    \begin{tabular}{c c c c c}
    \textbf{Data source} & \textbf{Total} & \textbf{Exp.} & \textbf{Det.} & \textbf{Acc.} \\
    \hline
    NSS ORBT    &    384 &    384 &  289 &$\sim$75\% \T\\
    NSS ASB1    &   1332 &   1332 & 1250 &$\sim$94\%\\
    CARMENES    &     48 &     48 &   34 &$\sim$71\%\\
    Control     &   5000 &      0 &    58 &$\sim$99\%\B\\
    \hline
    \end{tabular}
    \tablefoot{Expected (Exp.) column refers to the number of objects from the given sample that meet the chosen detectability criterion. Detected (Det.) column refers to the number of stars in the corresponding sample for which the model has computed a probability $p(y\mathbf{=}1)>p_0$. Accuracy (Acc.) reports model accuracy (detected/expected) on the given dataset.}
    \end{table}    
    
\begin{figure}[h!]
    \includegraphics[width=0.95\textwidth]{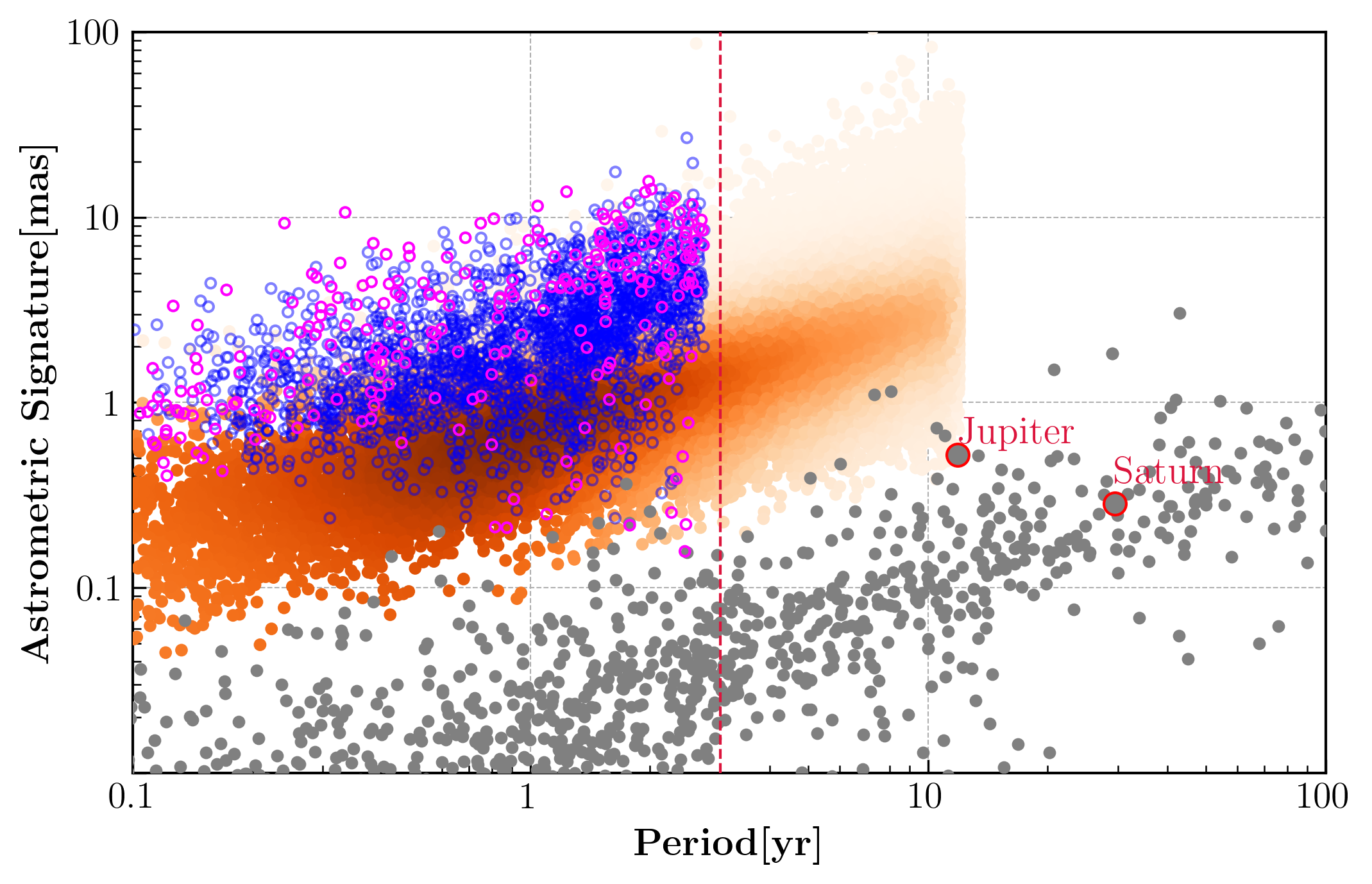}%
    \caption{Astrometric signature vs. orbital period for the simulated binary sources (orange), the 384 DR3 NSS ORBT solutions (magenta), the 1332 DR3 NSS ASB1 solutions (blue) and the currently confirmed exoplanets \citep{exoplanets_eu} (gray). The red dashed vertical line represents the DR3 time baseline of 34 months in comparison to the shorter periods of the Gaia DR3 unresolved binary orbital solutions. For the DR3 NSS solutions we used the expression $\alpha\mathbf{=}a_0/d$ to compute the astrometric signature. $a_0$ was computed using the Thiele-Innes elements ($A,B,F,G$) of the corresponding Gaia DR3 NSS solution and the recipe available in \citep[][Appendix A]{2023A&A...674A...9H}, and $d$ was computed as the inverse of the source parallax. For the simulated sources we used $\alpha\mathbf{=} M_2/M_1 \times a_1/d$ from  \cite{2014ApJ...797...14P}, where $M_1$ and $M_2$ are the primary and secondary masses, respectively, and $a_1$ is the semimajor axis of the orbit of the secondary.}
    \label{fig:astrometric_signatures}
    \end{figure}

    Figure~\ref{fig:astrometric_signatures} illustrates how the astrometric signature levels induced by the binary systems in our simulation are (intentionally) comparable to the signatures induced unresolved binaries so far detected by Gaia, and typically higher than currently confirmed exoplanets. Although the overall prediction results are not as optimal as we would like across all real samples, the performance of a model will always decrease when presented with real data. In particular, the characteristic DR3 calibration issues mentioned in Section~\ref{sect:agis_statistics} are more likely to affect the fainter and redder M-dwarfs from CARMENES sample. We therefore settled for a version of the model that showed a good compromise between the true and false positive rate, and decided to apply specific post-processing to tackle (at least partially) the mentioned calibration issues. We detail this approach in later sections.
    
    \section{Application to Gaia DR3}

    \subsection{ExoDNN application}\label{sect:model_application}
\begin{figure}[h!]
    \vspace{10pt}
    \includegraphics[width=9cm]{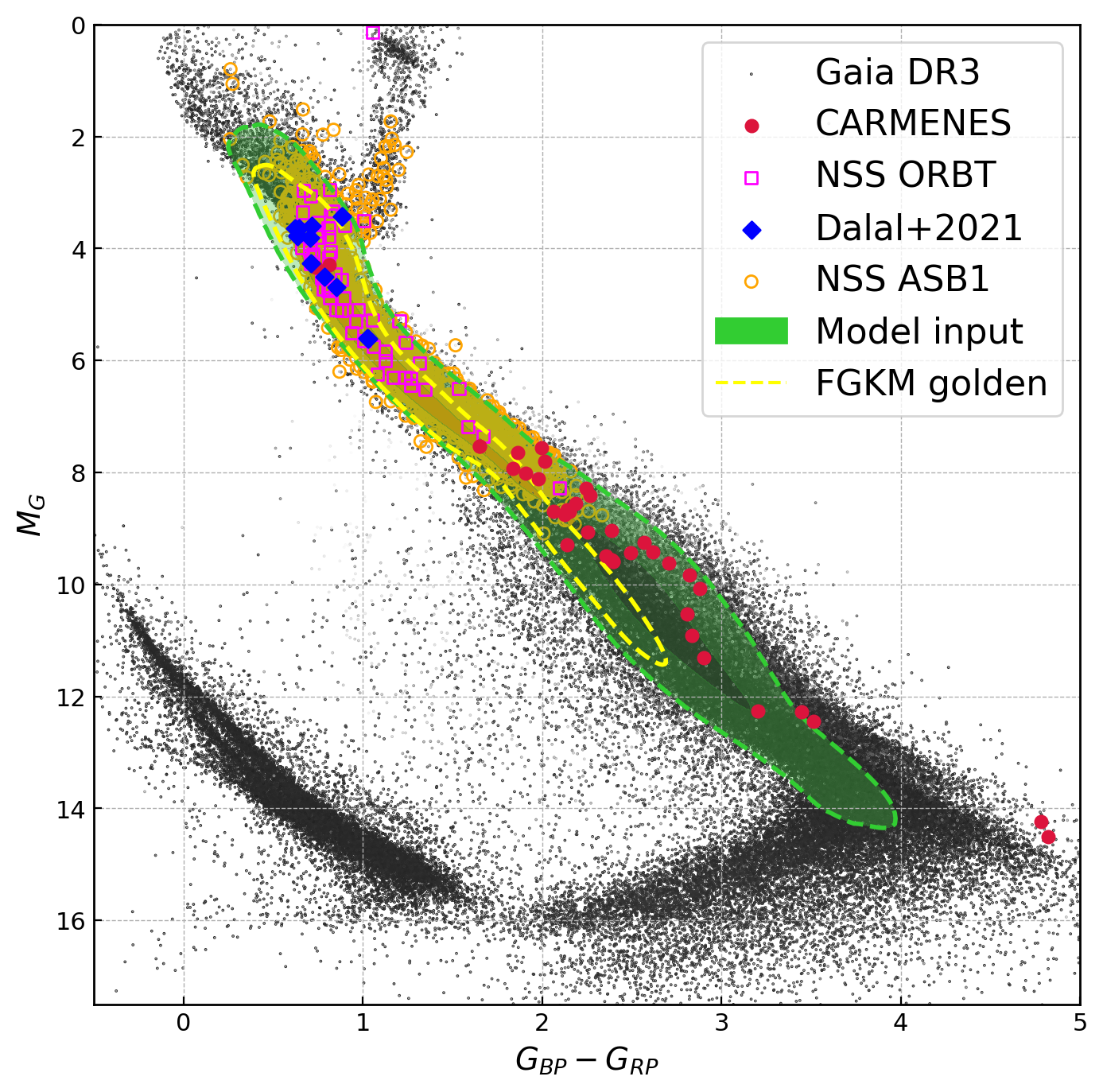}%
    \caption{Gaia $BP$ minus $RP$ color vs. absolute magnitude for the model input stars (green region). The FGKM golden sample is shown in yellow. Absolute magnitude is computed as: $m_G=phot\_g\_mean\_mag + 5.0 \cdot log(parallax)- 10$. Close binary systems where the smaller companion is a brown dwarf or a giant planet are displayed as diamonds (blue) and NSS Orbital* solutions are displayed as squares (magenta). CARMENES sources are displayed as circles (orange). Sources in correspond to the Gaia Catalog of Nearby Stars within $d<100pc$. }
    \label{fig:model_input_hrd}
    \end{figure}

    With our trained and tested deep learning model, we  turned to making predictions over real stars from Gaia DR3. We selected a volume limited sample of astrometrically well observed main sequence and giant stars at moderate distances ($d\mathbf{<}100pc$), from the Gaia DR3 Astrophysical Parameters sample \citep[Apsis,][]{2023A&A...674A..26C}. This dataset provides estimates for key stellar parameters such as mass, effective temperature and metallicity (among others), which were necessary to characterize the host stars of our potential candidates. To this sample, we applied a set of filtering criteria similar to those used for obtaining the FGKM golden sample of astrophysical parameters \citep[CR23 hereafter]{2023A&A...674A..39G}, but relaxed to reduce the heavy filtering of M stars that takes place in the golden sample. As such, our input for model prediction, fully contains the DR3 FGKM golden sample as a subset, but has a higher fraction of M stars in comparison (see Figure~\ref{fig:model_input_hrd}). 
    Full details on this process are provided in Appendix~\ref{appendix:model_input_data} and omitted here for brevity. 
    
    This selection resulted in a total of 102\,117 stars with estimated spectral types F, G, K, and M within 100pc to which we applied the same pre-processing steps used for the training data, namely the same robust scaling of the model attributes that was described in Section~\ref{sect:model_training}. The mentioned spectral types correspond to the \texttt{spectraltype\_esphs} parameter of the DR3 astrophysical parameters, which is computed using the low-resolution spectra from the BP-RP instrument onboard Gaia. We then used ExoDNN to compute the probability of each of these stars to host unseen companions. Our model returned 14\,606 stars with a predicted probability $p\! >\! p_0(=0.242)$, which where considered initial candidates. The breakdown per spectral type (\texttt{spectraltype\_esphs}) of these initial candidates is: 6017 M-type stars, 4180 K-type star, 2311 G-type stars and  2098 F-type stars.
    
    \subsection{Post-processing}\label{sect:additional_filter}
    We then applied a series of post-processing steps over ExoDNN prediction. These were aimed at obtaining a list of consolidated candidates sources to host unresolved companions. 
    \begin{figure*}[!h]
    \centering
    \includegraphics[width=5.7cm]{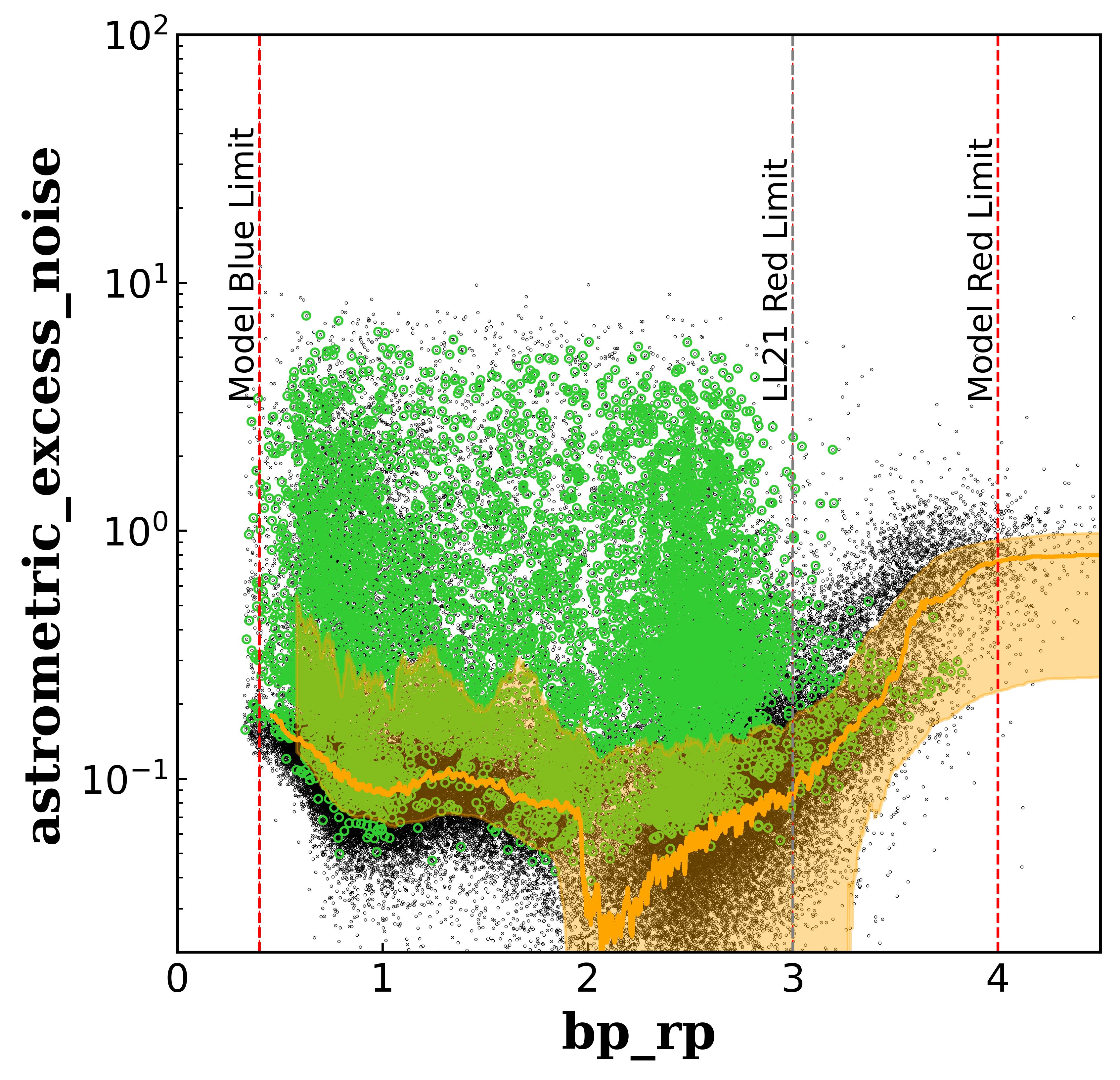}
    \hspace{5pt}
    \includegraphics[width=5.7cm]{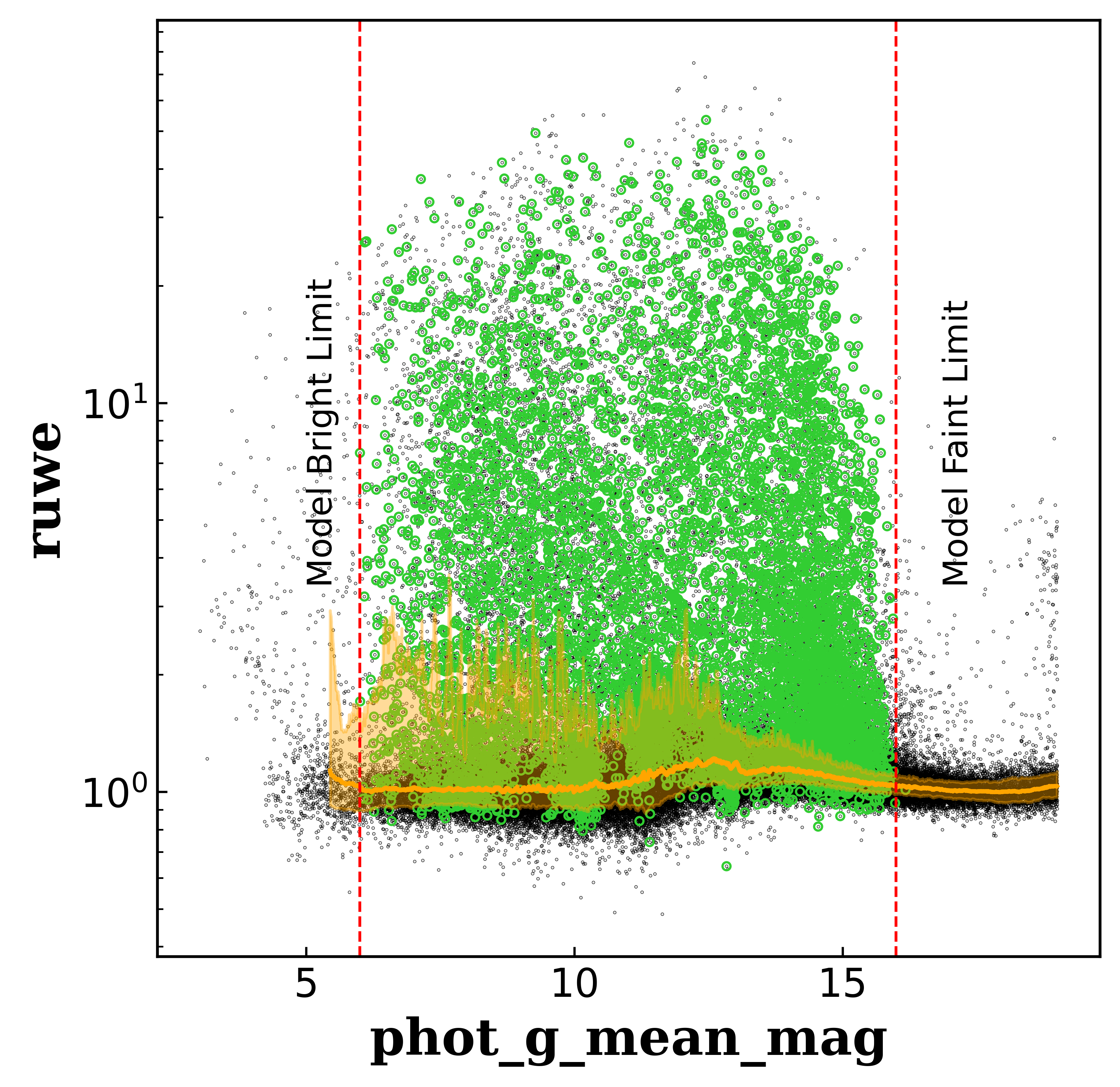}
    \hspace{5pt}
    \includegraphics[width=5.7cm]{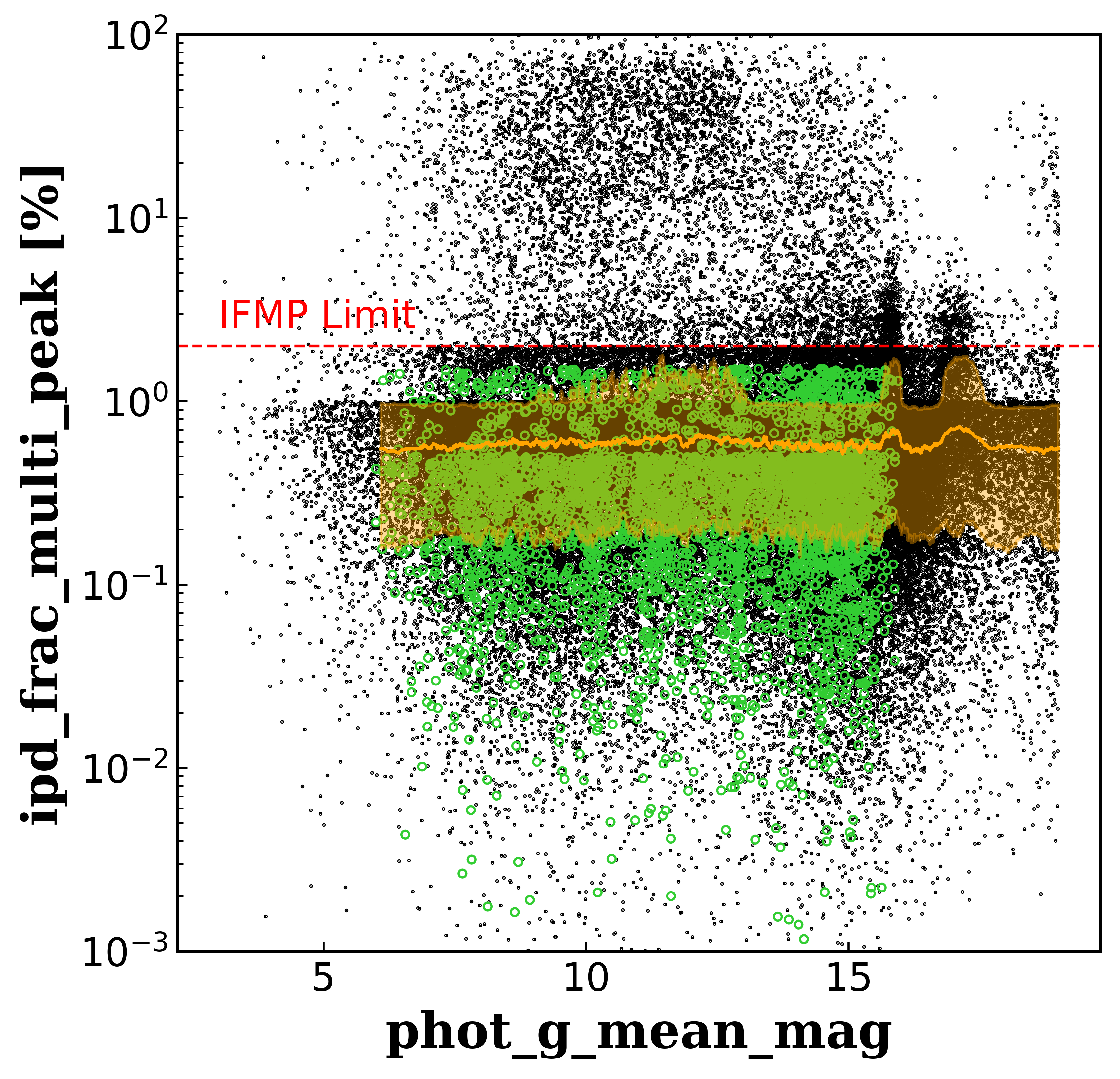}
    \caption{LEFT: \texttt{astrometric\_excess\_noise} vs. Gaia color \texttt{bp\_rp}. Model inputs are displayed in black and candidates in green. The vertical red lines show the applied color cuts. Orange line corresponds to the running median of the excess noise and the shaded orange region corresponds to the 16th-84th percentiles.MIDDLE: Equivalent plot but for \texttt{ruwe} vs. Gaia $G$ band magnitude. The vertical red lines show the applied bright and faint Gaia magnitude cuts. RIGHT: Equivalent plot but for \texttt{ipd\_frac\_multi\_peak} vs. Gaia $G$ magnitude. The horizontal red line corresponds to the selected cut-off limit for this parameter.}
    \label{fig:post_procesing}
    \end{figure*}
    First, we identified candidates that are already known binaries or with confirmed companions by doing a cross-match with a cone search radius of 1.8 arc-seconds against four different catalogues, the 9th Catalog of Spectroscopic Binaries \citep{2004A&A...424..727P}, the subset of Gaia DR3 NSS orbital solutions \citep{2023A&A...674A..34G} within 100pc, the Washington Double Stars catalogue \citep{WDS_2001AJ....122.3466M} and the online Encyclopedia of Exoplanetary Systems\footnote{Accessed early May 2025 at: \url{https://exoplanet.eu/home/}}. This cross-match identified 4383 stars which are present in either one or several of these catalogues. Some illustrative examples of known binaries that ExoDNN model has correctly identified among the initial list of candidates, are 61 Cyg (WDS J21069+3845AB), one of the closest binary systems, Struve 2398 (WDS J18428+5938AB), composed of two M-dwarfs, $\eta$ Cas (WDS J00491+5749AB), or the spectroscopic binaries 70 Oph (SBC9 1022) and HD137763 (SBC9 1638). Since the model is trained on binary systems without any internal hierarchical structure, it cannot distinguish between binaries and triples or quadruples. Therefore, systems such as $\kappa$ Tuc (WDS J01158-6853) or $\sigma$ CrB (WDS 16147+3352), were Gaia has resolved the components, are also reported as candidates by ExoDNN. Other prominent examples of stars within the prediction range of 100pc with confirmed substellar companions that the model correctly reports as candidates are: $\epsilon$ Eri, a young K-dwarf at $\sim$3pc with a confirmed $\sim 1.6M\text{Jup}$ companion \citep{2006AJ....132.2206B}, 
    TZAri, a close M5.0V star located at $\sim$4.8pc that hosts an RV detected Saturn-type exoplanet ($\sim0.21M_\text{Jup}$) \citep{2022A&A...663A..48Q}, $\upsilon$ And at $\sim$13.5pc orbited by several exoplanets including a Jupiter-like companion detected through RV \citep{2011A&A...525A..78C}, and BD+05 5218 (HIP 117179) a G3V-dwarf 88pc away that hosts an astrometrically detected $\sim$44$M_\text{Jup}$ brown dwarf \citep{2023MNRAS.526.5155S}. 
    
    Second, we tackled the calibration issues affecting AGIS statistics (Section~\ref{sect:agis_statistics}), by following \cite{2021A&A...649A...2L}, Section 5.3 recommendations. We removed sources with $\texttt{astrometric\_excess\_noise\_sig} < 2$, and with colors ($G_{BP}-G_{RP}\mathbf{\leq}0.4$), or ($G_{BP}-G_{RP}\mathbf{\geq}4.0$). We extended the red end color limit to avoid filtering out known systems which our model correctly considers candidates and would otherwise be filtered out using the recommended constrain of ($G_{BP}-G_{RP}>3.0$). We also removed sources with ($\texttt{phot\_g\_mean\_mag} < 6$) and ($\texttt{phot\_g\_mean\_mag} > 16$), as \texttt{ruwe} can be unreliable beyond those magnitude. This filter removed 1579 sources.
    
    Third and final, we used the \texttt{ipd\_frac\_multi\_peak} parameter as an additional discriminator. This statistic is sensitive to the presence of multiple point-spread-functions (PSF) within the same detection window. This situation can occur depending on the scan direction when a partially resolved binary is observed by Gaia. We adopted the suggested value for filtering on this statistic from \citep{2021A&A...649A...5F}, and only retained sources with \texttt{ipd\_frac\_multi\_peak}<2. Because this parameter is stored as an integer percent, a value of 2 actually means that 0–1\% of the transits detected for a given source were multi-peaked. Hence, this filter aims to obtain clean single-peaked sources were the chances of any astrometric perturbation due to a close (partially unresolved) companion of stellar nature is close to zero. This filter removed 1230 sources.
   
    In Figure \ref{fig:post_procesing} we illustrate (from left to right) the different cuts in color, magnitude and fraction of multiple PSF found in the detection window, applied on each successive post-processing filter to obtain the final candidates. All together, these three post-processing steps removed 7\,192 stars from the original candidate list, resulting in a total of 7414 stars which were considered final candidates. The breakdown per spectral type of the final 7414 candidates is: 3757 M-type stars, 1852 K-type star, 950 G-type stars, and 855 F-type stars. 

    \subsection{List of candidates}

\begin{figure*}[!h]
    \includegraphics[width=12.5cm]{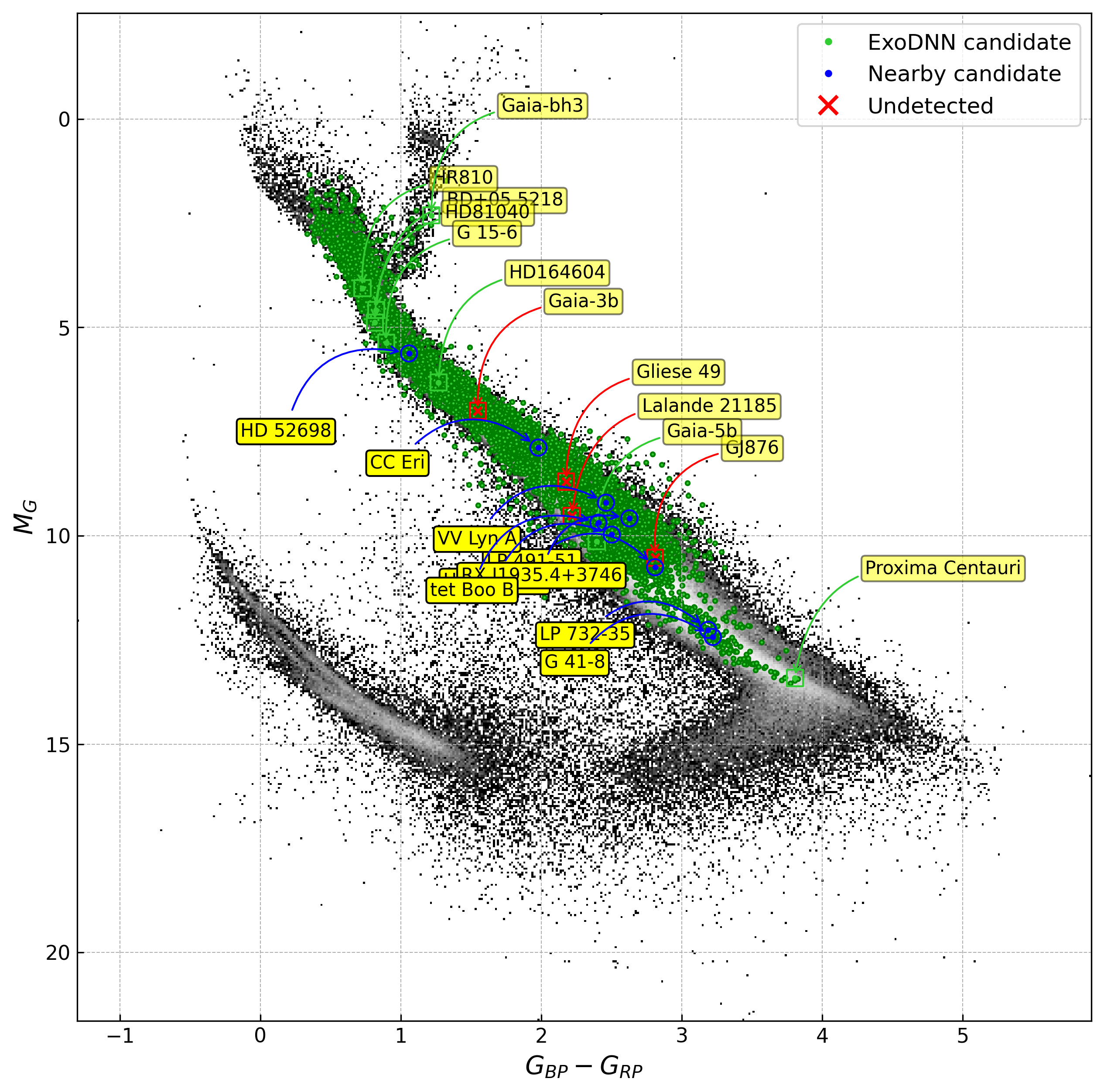} 
    \caption{Gaia BP minus RP color vs. absolute magnitude diagram of Gaia DR3 Catalog of Nearby Sources. The 7414 final candidates reported by ExoDNN after filtering are displayed in green. Prediction for a few reference objects, Proxima Centauri, LHS1610, BD+05\;5128, HD164604, Gaia-3b, Gaia-4b and Gaia-5b, etc., is also shown. Sources with known companions detected by ExoDNN are marked by the green boxes, while red-crosses refer to reference objects for which model predicted probability $\hat{p}<p_0$, that is, undetected. ExoDNN closest new candidates are highlighted to the left of the main sequence with blue circles.}
    \label{fig:candidates_hr_diagram}
    \end{figure*}

\begin{figure}[!h]
    \includegraphics[width=8cm]{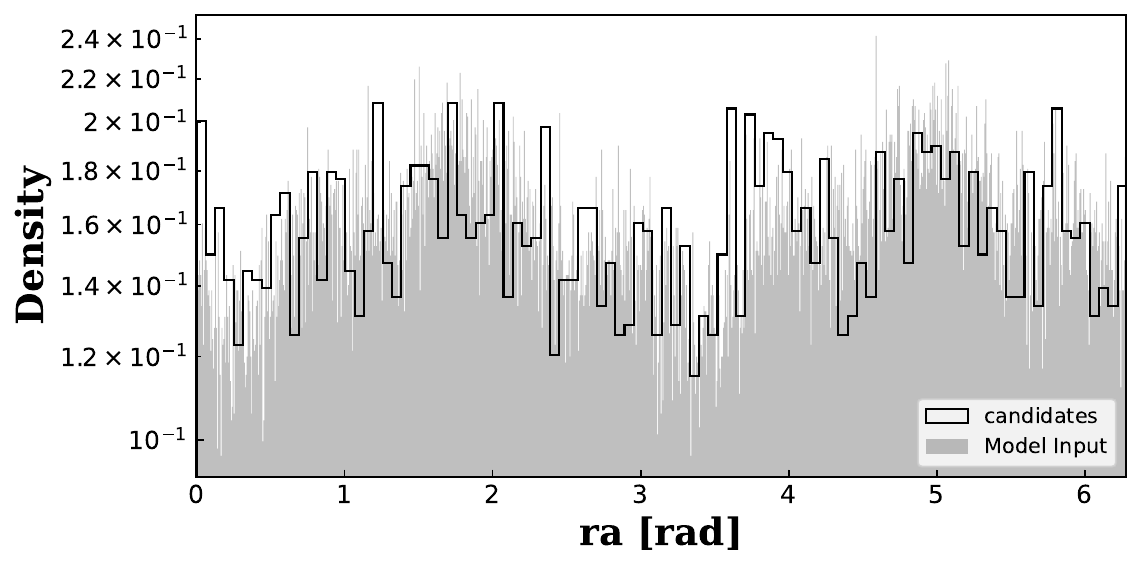}
    \vspace{10pt}
    \includegraphics[width=8cm]{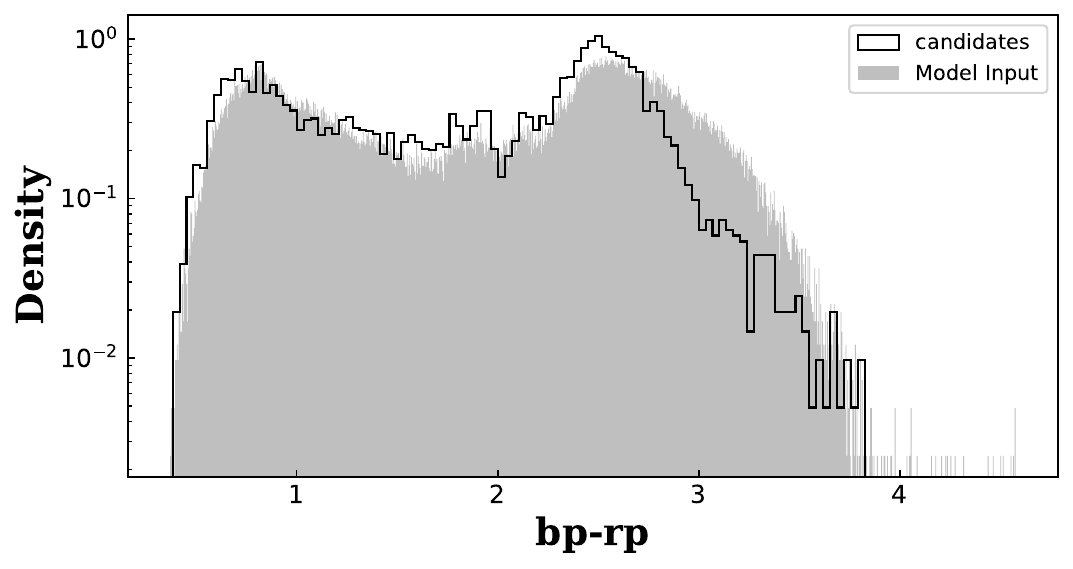}
    \vspace{10pt}
    \includegraphics[width=8cm]{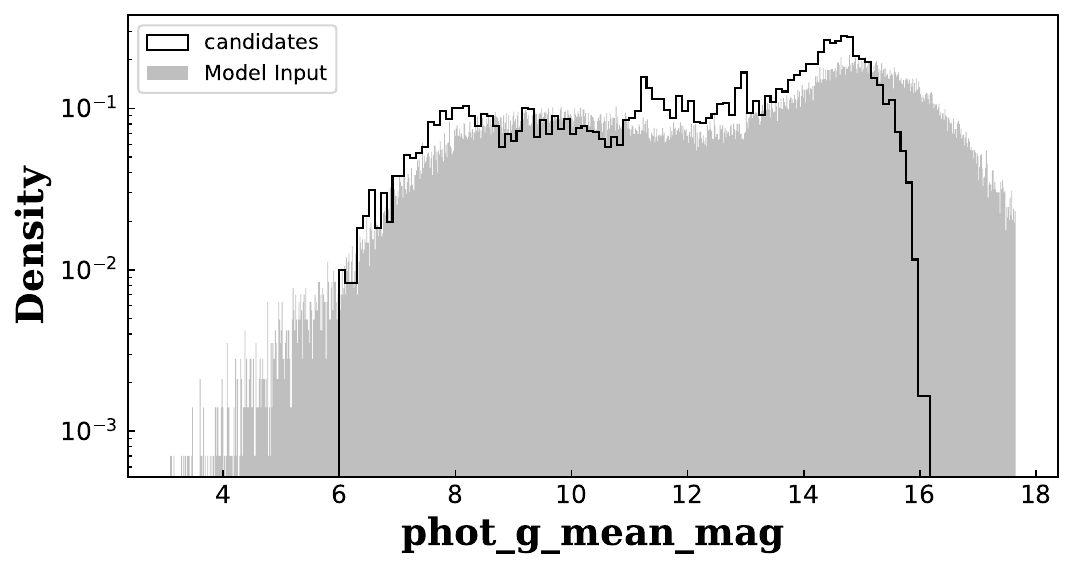}
    \caption{TOP: Distribution of the right ascension values for candidates sources (solid black line) and the original input for the model (gray) MIDDLE: Equivalent display but for Gaia BP minus RP color (\texttt{bp-rp}) distribution. BOTTOM: Equivalent display but for the Gaia magnitude (\texttt{phot\_g\_mean\_mag}) in this case.}
    \label{fig:candidates_check}
    \end{figure}

    We show in Figure~\ref{fig:candidates_hr_diagram} a color-magnitude diagram of the Gaia Catalog of Nearby Stars \citep{2021A&A...649A...6G}, where we highlight the final 7414 consolidated candidates reported by ExoDNN, along with a few sources to help showcase ExoDNN predictions.
    These showcase examples are Gaia-4b and Gaia-5b \citep{2025AJ....169..107S}, a massive planet and a brown dwarf, respectively, with masses of $m\mathbf{=}11.8\pm\,0.7M_\text{Jup}$ and $m\mathbf{=}20.9\pm\,0.5M_\text{Jup}$, the only substellar companions from DR3 that have been externally confirmed so far via RV follow-up observations. Both Gaia-4b and Gaia-5b are reported as candidates by ExoDNN. G 15-6, where \cite{2025MNRAS.537.1130S} have recently identified a brown dwarf companion with an astrometric mass $m\mathbf{=}62.4\pm{6.2}M_\text{Jup}$, is also reported as a candidate. We also show Gaia black hole 3 \citep{2024A&A...686L...2G}, whose host star is also detected as a candidate. We omit Gaia-1b and Gaia-2b since these are close-in transiting planets \citep{2022A&A...663A.101P} with very short orbital periods ($\sim$3 days) that do not produce enough gravitational pull to be astrometrically detectable in DR3 (or even DR4/5). Another interesting example is HD164604, a K-dwarf at $\sim$40pc orbited by a $\sim$14$M_\text{Jup}$ super Jupiter and detected via astrometry \citep{2023A&A...674A..34G}, that is also reported as a candidate by ExoDNN. We searched for signs of preferential detection of candidates on particular sky positions, distances, Gaia colors, or Gaia magnitudes. Figure~\ref{fig:candidates_check} shows the distributions of some of those parameters, which do not reveal any significant bias in the model output candidates.   
    When we analyse the nearby candidates reported by ExoDNN, shown in Table~\ref{tab:top_10_candidates}, we can see that the majority are M-dwarfs. This is not totally surprising since the astrometric signature of an hypothetical companion will be stronger in an M-dwarf due to the higher mass ratio. This, together with their abundant numbers ($\mathbf{\sim}75\%$ of the stellar population \citealt{2024ARA&A..62..593H}), makes M stars usual targets for RV exoplanet surveys \citep[e.g.,][]{2023A&A...670A.139R}. We find among these closest candidates an illustrative example (HD 52698) on how the model decision is driven by the most influential parameters \texttt{ruwe} and \texttt{astrometric\_excess\_noise}. This spectroscopic binary has intrinsic high values of both astrometric fit statistics, which drives ExoDNN to make a strong prediction of the probability that this source has companions, 0.99 probability against the average of the other 12 candidates 0.37.
    
    To check whether any biases had been introduced by adopting a deep learning approach in this domain, we compared ExoDNN predictions with those from \cite{2025MNRAS.537.1130S}. In that work, the authors used a semisupervised anomaly detection framework that combines feature importance and ensemble classifiers to identify potential companions to stars. The scientific goal (astrometric detection of substellar companions) is equivalent to ours, but the technique is different, thus making the comparison relevant. We fed ExoDNN with the 22 high-probability candidate sources reported by these authors (see their Table 4). To do this, we preprocessed the required model parameters corresponding to each of those 22 Gaia DR3 sources and then computed the probability of each of them to host a substellar companion using ExoDNN. We present the predictions made by ExoDNN over those 22 candidates in Table \ref{tab:22_sahlmann_candidates}. ExoDNN predicts that 16 out of 22 ($\mathbf{\sim}$72\%) also are candidates to host companions. However, as noted by \cite{2025MNRAS.537.1130S} (see their section 4.2), some of these candidates (e.g., * 54 Cas and BD+75 510), were identified after DR3 release as false positives. Finally, the six sources not considered as candidates by ExoDNN have, in relative comparison to the others, a combination of \texttt{ruwe} and \texttt{astrometric\_excess\_noise} that results in a lower probability than our established threshold. The rest of the candidates (16) are detected by ExoDNN with computed probabilities to hold companions in the range (0.25 to 0.99) depending on the relative values of the two main astrometric fit quality indicators. We find cases such as HD 207740, HD 40503, GSC 09436-01089 and G15-6, where the relative high values of both the \texttt{ruwe} and \texttt{astrometric\_excess\_noise} drive ExoDNN prediction to a high predicted probability of holding companions. This result is reassuring, as it represents an independent check of the ExoDNN behavior.

    \begin{table*}[ht]
    \caption{ExoDNN closest-candidates prediction.}
    \label{tab:top_10_candidates}
    \small
    \centering
    \begin{tabular}{cccccccc}
    \hline
    \T\textbf{Name} & \textbf{Alt. Name} & \textbf{Gaia designation} & \textbf{Parallax}[mas] & \textbf{Sp Type} & \textbf{ruwe} & \textbf{$\epsilon$}[mas] & \textbf{$\hat{p}$}\T\\
    \hline
    \T G 158-27    & GJ 1002  & Gaia DR3 2441630500517079808 & 206.35 & M5.5V$^{7}$  & 1.34 & 0.36 & 0.29 \\
    CD Cet          & GJ 1057  & Gaia DR3 3179036008830848  & 116.28  & M4.5V$^{8}$  & 1.45 & 0.35 & 0.27 \\
    Wolf 1069   & GJ 1253  & Gaia DR3 2188318517720321664   & 104.44  & dM5.0$^{6}$  & 1.42 & 0.31 & 0.33 \\
    LP 732-35   &  GJ 3668 & Gaia DR3 3560746351897773952   & 86.90   & M4.5V$^{1}$  & 1.90 & 0.36 & 0.52\\
    CC Eri      &  GJ 103  & Gaia DR3 4946938113149426944   & 86.61   & K7.V C$^{10}$& 2.26 & 0.38 & 0.33\\
    G  41-8     &  --      & Gaia DR3 605079910398496640    & 85.80   & M6.0V$^{2}$  & 1.12 & 0.20 & 0.35\\
    VV Lyn A    &  GJ 277A & Gaia DR3 898506093171335424    & 83.38   & M4.5V$^{1}$  & 2.51 & 0.41 & 0.37\\
    HD 304043   &  GJ 442  & Gaia DR3 5339892367684811520   & 78.89   & M3.5V$^{1}$  & 1.01 & 0.34 & 0.31\\
    RX J1935.4+3746 & --   & Gaia DR3 2051984436005124480   & 69.12   & M3.0V$^{3}$  & 2.20 & 0.42 & 0.50\\
    $\theta$ Boo B &  GJ 549 B   & Gaia DR3 1604859408265509632& 68.81& M2.5V$^{4}$  & 1.15 & 0.35 & 0.36\\
    G 268-38    & GJ 3053 & Gaia DR3 2371032916186181760    & 66.83   & M4.5V$^{1}$  & 1.53  & 0.28 & 0.29 \\     
    HD 52698    &  GJ 259 & Gaia DR3 2920772722738017920    & 67.80   & K1.V B$^{5}$ & 13.03 & 2.16 & 0.99\\
    LP 491-51   &  --     & Gaia DR3 3968118399284369280    & 65.84   & M3.0V$^{3}$  & 3.17  & 0.54 & 0.43\B\\
    \hline
    \end{tabular}
    \tablefoot{$\texttt{ruwe}$ and \texttt{astrometric\_excess\_noise} $\epsilon$ values are from Gaia DR3 main source table. The column $\hat{p}$, is the probability computed by ExoDNN that the given source hosts a detectable companion. (1) \citealt{1995AJ....110.1838R};(2) \citealt{2014AJ....147...20N};(3) \citealt{2009ApJ...699..649S};(4) \citealt{2015A&A...577A.128A};(5) \citealt{2006A&A...460..695T};(6) \citealt{2019A&A...625A..68S};(7) \citealt{2024A&A...684A...9S};(8)\citealt{2013AJ....145..102L};(9) \citealt{1989ApJS...71..245K};(10) \citealt{2006A&A...460..695T}.}
    \end{table*}
    
    \begin{table*}[ht]
    \caption{ExoDNN prediction on external candidates.}
    \label{tab:22_sahlmann_candidates}
    \small
    \begin{tabular}{ccccccc}
    \hline
    \T\textbf{Name} & \textbf{Alt. name} & \textbf{Gaia designation} & \textbf{Parallax}[mas] & \textbf{ruwe} & \textbf{$\epsilon$}[mas] & \textbf{$\hat{p}$}\T\\
    \hline
    \T HD 207740 &      HIP 107821 & Gaia DR3 1576108450508750208 & 14.020275 & 5.338995 & 1.204246 & 0.99 \\
    HD 40503 & HIP 28193 & Gaia DR3 5323844651848467968 & 11.477482 & 5.047571 & 0.909602 & 0.99 \\
    GSC 09436-01089 & - & Gaia DR3 3921176983720146560 & 15.452359 & 3.036620 & 0.617942 & 0.95 \\
    G15-6 & HIP 73800 &  Gaia DR3 1897143408911208832 & 20.304506 & 3.340308 & 0.644446 & 0.94 \\
    HD 206484 & - & Gaia DR3 2171489736355655680 & 13.936007 & 2.631042 & 0.536911 & 0.79 \\
    HD 49264 & HIP 32253 & Gaia DR3 3913728032959687424 & 14.752394 & 2.584024 & 0.482179 & 0.75 \\
    HD 212620 & BD +75 825 & Gaia DR3 5148853253106611200 & 13.800585 & 2.746541 & 0.417314 & 0.75 \\
    HD 106888 & HIP 59933 & Gaia DR3 1156378820136922880 & 13.872044 & 3.363915 & 0.353119 & 0.67 \\
    BD+3292 &   HIP 2666 & Gaia DR3 2540855308890440064 & 10.091830 & 2.012354 & 0.272028 & 0.51 \\
    HD 128717 & BD +58 1514 &  Gaia DR3 4545802186476906880 & 18.444739 & 2.028104 & 0.315474 & 0.40 \\
    HD 104289 & BD +60 1368 & Gaia DR3 364792020789523584 & 12.960809 & 1.581556 & 0.272007 & 0.36 \\
    TYC 4998-437-1 & - & Gaia DR3 3909531609393458688 & 4.687014 & 1.720975 & 0.276329 & 0.36 \\
    LP 769-9 & - &  Gaia DR3 1878822452815621120 & 16.052830 & 1.917616 & 0.236748 & 0.34 \\
    * 54 Cas &  HIP 10031 & Gaia DR3 5484481960625470336 & 8.776449 & 1.982503 & 0.246423 & 0.30 \\
    BD-06 2423A & - & Gaia DR3 6330529666839726592 & 7.563384 & 1.626470 & 0.193307 & 0.30 \\
    BD+75510 & BD +75 510 & Gaia DR3 5773484949857279104 & 4.225611 & 1.437869 & 0.176325 & 0.25 \\
    HD 153376 & BD +15 3089 & Gaia DR3 2280560705703031552 & 7.321575 & 1.419174 & 0.193269 & 0.21 \\
    HD 206848 & BD +23 4384  & Gaia DR3 1610837178107032192 & 13.560875 & 1.320370 & 0.214271 & 0.21 \\
    TYC 277-599-1 & - & Gaia DR3 2884087104955208064 & 25.506822 & 1.408193 & 0.156793 & 0.13 \\
    HD 78631 & CD-51 3504 & Gaia DR3 522135261462534528 & 37.011715 & 0.876396 & 0.112947 & 0.12 \\
    HD 99251 & BD+08 2498 & Gaia DR3 1712614124767394816 & 28.249344 & 1.019851 & 0.083416 & 0.11 \\
    BD+24 4592 & HIP 110894 & Gaia DR3 3067074530201582336 & 14.480729 & 1.580362 & 0.170151 & 0.05 \B\\
    \hline
    \end{tabular}
    \tablefoot{$\texttt{ruwe}$ and \texttt{astrometric\_excess\_noise} $\epsilon$ values are from Gaia DR3 main source table. The column $\hat{p}$, is the probability computed by ExoDNN that the given source hosts a detectable companion.}
    \end{table*}
    
    \section{Discussion}\label{sect:results}
    \subsection{Effects of filtering}
    
\begin{figure}[h!]
    \includegraphics[width=9cm]{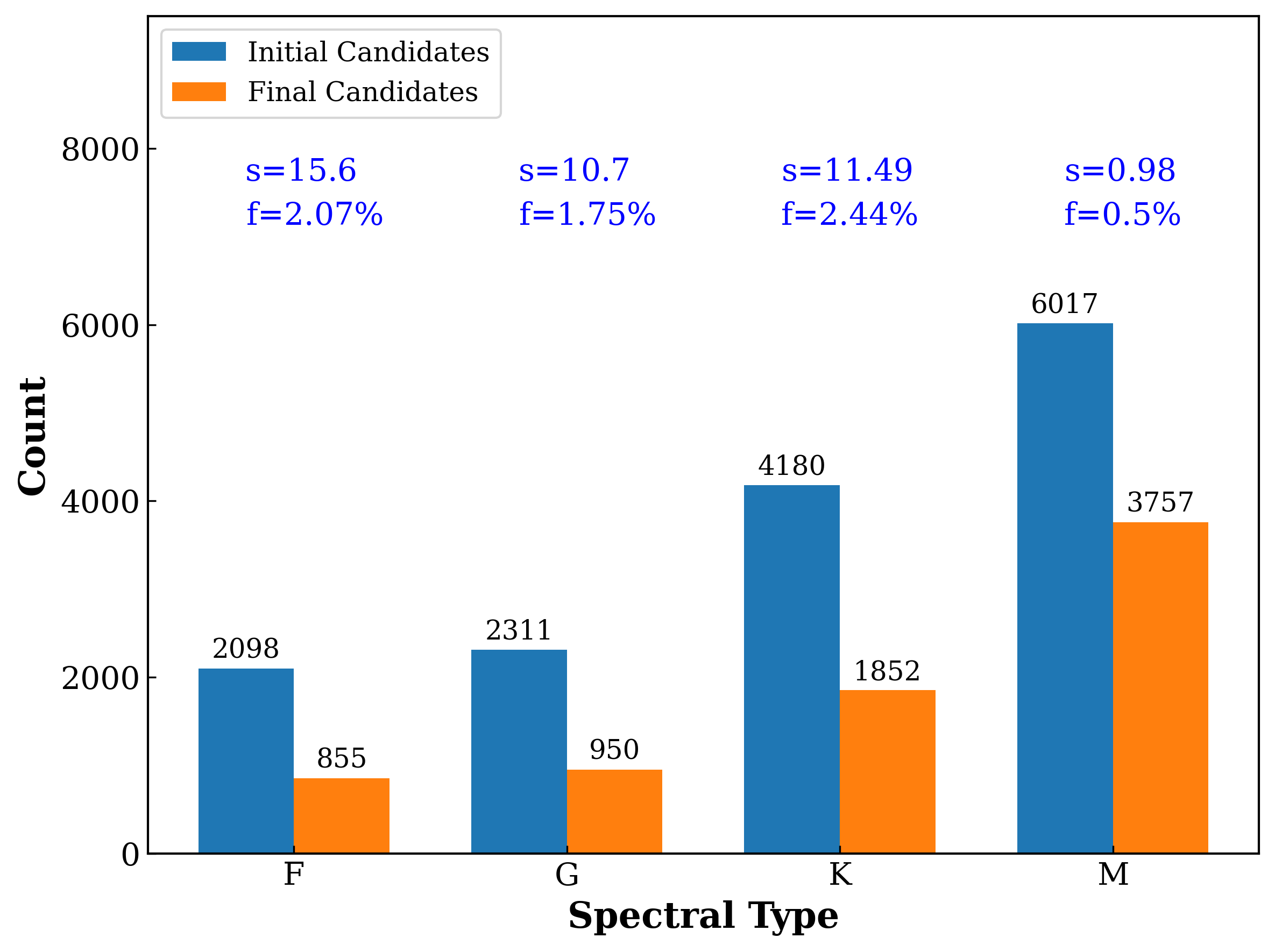}%
    \caption{Overall number of candidates per spectral type (\texttt{spectraltype\_esphs}), pre-filtering (blue) and post-filtering (orange). Blue text labels on top of each group of bars correspond respectively to ($s$): median values of the \texttt{astrometric\_excess\_noise\_sig} parameter for the given spectral type and ($f$): median values of the  \texttt{ipd\_frac\_multi\_peak} parameter for the given spectral type.}
    \label{fig:candidates_breakdown}
    \end{figure}

    The significant reduction ($\sim$50\%) from the original 14\,606 candidates to the final 7414 has its origin in the post-processing filters we applied (Section \ref{sect:additional_filter}). Over 30\% of the original candidates were identified during the cross-match (first filter) as previously known binary or multiple systems. These candidates are considered  true positives, which ExoDNN has correctly identified and thus, they were removed because the focus of our study is on finding genuinely new candidates. The remaining filtering steps remove 20\% of the initial candidates but affect differently across the spectral types. While the astrometric excess noise significance filter (second filter) affects  the fainter (M-type) stars more heavily, the multi peak discriminator (third filter) affects more F, G and K-type stars. The net effect of these filters in the total numbers per spectral type is illustrated in Figure~\ref{fig:candidates_breakdown}.
    \begin{figure}[!h]
    \includegraphics[width=8.5cm]{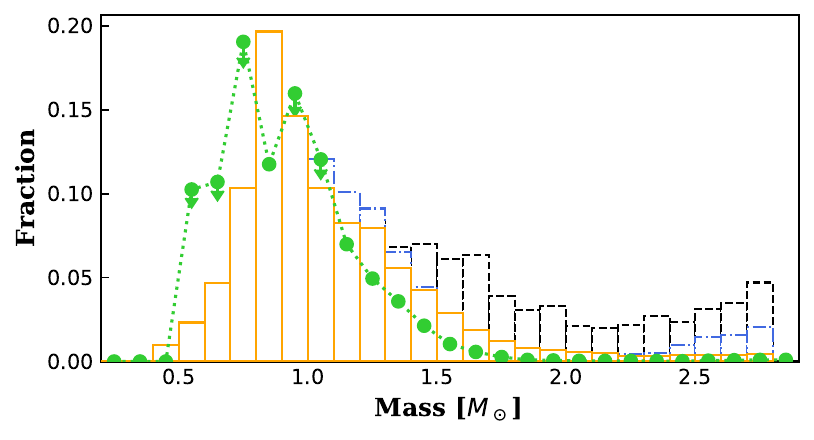}
    \vspace{5pt}
    \includegraphics[width=8.5cm]{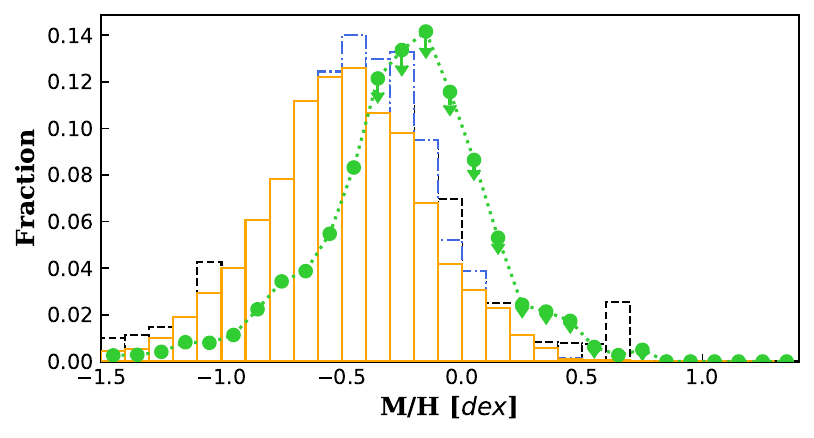}
    \caption{LEFT: Primary mass distribution of the candidate stars reported by ExoDNN (green). Also added for comparison are the mass distributions corresponding to the NSS Orbital* (orange), AstroSpectroSB1 (blue), and Eclipsing* binary (black) solutions from the DR3 NSS. The ExoDNN predictions show the 6\% model false positive rate as downward arrows. RIGHT: Equivalent distributions but for metallicity of the candidates reported by ExoDNN.}
    \label{fig:candidates_mass_comparison}
    \end{figure} 
    
    \noindent Another prominent feature of our final candidates, is the apparent depletion of faint candidates ($M_G\mathbf{>}12$). This is due to the strong filtering induced by the second post-processing step (Section \ref{sect:additional_filter}) for faint magnitudes. The fraction of DR3 sources with significant ($\mathbf{>}2$) excess noise decreases from 97\% at $9\mathbf{\le}G\mathbf{\le}12$ to only 20\% at $G\mathbf{=}14$ \citep[][Table 4]{2021A&A...649A...2L}, and therefore that filter is particularly efficient in removal of M-type candidates in the magnitude range ($12\mathbf{\le}G\mathbf{\le}15$). These sources frequently exhibit excess noise significance values lower than 2, see Figure \ref{fig:candidates_breakdown} ($s$) values. Also, the significant depletion of $\mathbf{\sim}50\%$ F and G-type initial candidates is caused by our third filtering step. The general increased multiplicity fraction of F and G dwarfs, $\sim$60\% for F-G stars vs. $25\mathbf{-}40\%$ for M-dwarfs \citep{2013ARA&A..51..269D}, will result in relatively higher numbers of the \texttt{ipd\_frac\_multi\_peak} parameters for the spectral types F, G and K over M-type sources, see Figure \ref{fig:candidates_breakdown} ($f$) values. Therefore the third filtering step removes primarily the brighter and more massive initial candidates. We plan to improve these features in further versions of ExoDNN by applying refined selection criteria for sources around the lower astrometric excess noise significant limit and for those with multiple psfs within the detection window.
   
    \subsection{Host characterization}
    
    Thanks to the selection of the model input from the DR3 Astrophysical Parameters dataset, we have access to stellar parameters such as the mass, metallicity, and age of the candidate stars. These parameters have been derived by the astrophysical parameter inference system \citep[Apsis,][]{2023A&A...674A..28F}, using the low-resolution spectra provided by Gaia's blue and red photometers (BP/RP), and provide valuable information to characterize our candidate stars population. If we analyse the corresponding distributions of mass, metallicity and age for the candidates reported by ExoDNN , we find a peak mass $M_\text{cand.}\mathbf{=}0.7 \pm 0.1 M_\odot$ and a peak metallicity of $M/H_\text{cand.}\mathbf{=}-0.1 \pm 0.01 \text{dex}$ (see Figure~\ref{fig:candidates_mass_comparison}). This would suggest that the candidates found by ExoDNN are similar, in terms of stellar properties, to the NSS astro-spectroscopic binaries, but not to the NSS eclipsing binaries (see \citealt{2023A&A...674A..34G}). The model has been trained on a distribution of periods that overlaps with the one from the NSS astro-spectroscopic binaries (typically hundreds of days), which results in a model that is more sensitive to those orbital configurations rather than to the much shorter orbital periods of the eclipsing binaries (typically a few days). 

    \noindent It should be noted that the parameters derived by the Apsis may suffer from calibration problems in the case of unresolved binaries \citep[Section 3.5.2]{2023A&A...674A..28F}. This is due to the lack of sufficiently high-quality synthetic models of BP and RP spectra of unresolved binaries and, therefore, our comparison can only be regarded as qualitative. 

    \subsection{Caveats}
    
    We remark that in its current version, ExoDNN reports candidate stars to host companions detectable by Gaia, but it cannot assess the nature (stellar or substellar) of the potential companions. In general, when the current ExoDNN reports a source as a candidate, we may face different scenarios. One, where the signatures of existing companions in the system are of the order of the DR3 measurement uncertainty. In this case, ExoDNN prediction may be just detecting the signatures of existing companions in the system or alternatively, indicating the presence of a yet unknown companion in a longer period orbit. Another possible scenario occurs when the source has no reported companions yet, but ExoDNN predicts their presence. And yet another scenario may occur, which represents an astrophysical false positive. In this latter case, the AGIS statistics are inflated but for reasons other than multiplicity, triggering a false positive. Irrespective of the case, supporting evidence such as additional RV measurements, epoch astrometry or imaging data, are necessary to first confirm or rule out the existence of companions, and only then be able to provide estimates on their masses.
    
    \section{Conclusions}
    Using the rich dataset provided in Gaia DR3, we created a deep learning model, ExoDNN, that uses the available DR3 parameters to predict the probability of a Gaia DR3 star to host one or more companions. To do this, ExoDNN takes advantage of the existing correlation between AGIS fit quality statistics and the deviation from a single source model. We have shown how ExoDNN is able to correctly detect different types of multiplicity, from known binaries on wide orbits to star+brown dwarf configurations. When applied to a sample of stars from the Gaia DR3 catalogue within 100pc, ExoDNN generated a list of 7414 new candidates to host unresolved companions. The number of candidate stars detected is comparable in order of magnitude to predictions from earlier studies \citep[e.g.,][]{2008A&A...482..699C,2014ApJ...797...14P}. The post-processing steps we applied are conservative,  which resulted in the removal of approximately $\sim$50\%  of the initial candidates found by ExoDNN, but this step was necessary to avoid contamination from sources possibly affected by DR3 AGIS calibration limitations. A false positive rate of $\mathbf{\sim}1.2\%$ is expected from the current version of ExoDNN and should be taken into account when assessing the proposed candidates by performing additional scrutiny on a per-source basis. The main limitation of the current version of ExoDNN is the inability to determine the nature (stellar or substellar mass) of potential companions without incorporating external data, such as RV measurements or epoch astrometry.
    \begin{figure}[!h]
    \includegraphics[width=8.5cm]{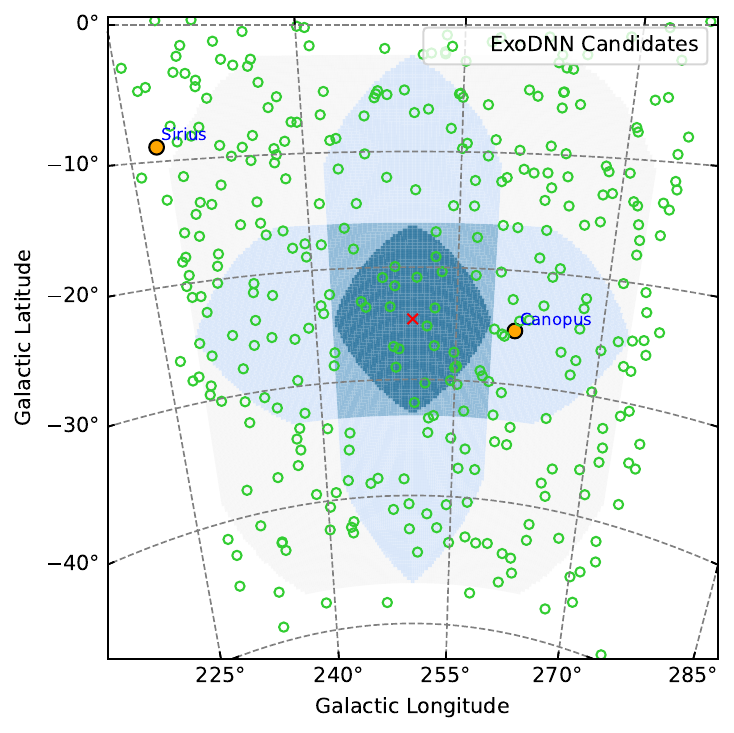}
    \caption{ExoDNN G-type star candidates overlay in PLATO first field of view (LOPS1).}
    \label{fig:candidates_on_platp_fov}
    \end{figure}
    In addition, since the ExoDNN predictions are generated from DR3 astrometry, those predictions are limited by the available precision in DR3 measurements, which has been the main driver in applying ExoDNN to volume-limited sample of stars ($d$<100pc). Technically speaking, however, nothing prevents applying the model to sources further away (one extreme example is ExoDNN prediction over Gaia BH3, located at $d\mathbf{\sim}590pc$). Despite these limitations and until Gaia DR4 is published along with a list of exoplanet candidates, ExoDNN candidates could be a powerful tool for the community for benchmarking different theories of planetary formation given its statistical relevance. In addition, the proposed solar-type (G) candidates could be interesting in the context of future  missions such PLATO \citep{2022EPSC...16..453R}. In Figure~\ref{fig:candidates_on_platp_fov}, we display the G-type candidates found by ExoDNN over the PLATO first field of view, FoV South 1 (240 G-dwarf candidates). 
    
    We plan to further enhance the current model performance in several ways. First, we would like to improve the model sensitivity specially over the faint end covering M-dwarfs. Training ExoDNN on additional examples of real M-type stars that are known to host companions could help it learn (at least partially), the intrinsic Gaia DR3 calibration artifacts affecting these sources. This, in turn, could eliminate the current need for applying specific excess-noise cuts to M-type candidates. Then, we will also revise the current model architecture to include uncertainty estimations. This will introduce the ability to inform the user on the uncertainty in a given probability prediction and require the evolution of the current neural network architecture towards a deep probabilistic architecture \citep[e.g.,][]{neal2012bayesian,blundell2015weight,lakshminarayanan2017simple,abdar2021review}. Last but not least, to gain the ability to asses the nature of  reported candidates, we will need to add external datasets, such as RV values or astrometric time series. This will allow us to constrain the candidate companion masses and increase the current scope of the model from astrometric detection of substellar companions to a broader one, namely, the characterization of substellar companions.

    Some of our candidates are close enough to make them suitable targets for follow-up observations by different means. For candidates at moderate distances, RV follow-up observations with high-resolution spectrographs such as HARPS \citep{2012SPIE.8446E..1VC} or ESPRESSO \citep{2021A&A...645A..96P}, would allow for the presence of companions to be confirmed or for orbital configuration and masses to be at least partially constrain. We intend to asses the suitability for follow-up observations of the most promising candidates reported in this work and identify the best follow-up strategy.

    \section*{Data availability}
    The full version of the 7414 new Gaia DR3 candidate stars to host companions is only available in electronic form at the CDS via anonymous ftp to cdsarc.u-strasbg.fr (130.79.128.5) or via \url{https://cdsarc.cds.unistra.fr/viz-bin/cat/J/A+A/704/A150}. 
    
    \newpage
    \begin{acknowledgements}
    C.C. acknowledges financial support by the Consejo Superior de Investigaciones Científicas (CSIC) through the internal project 2023AT003 associated to the RYC2021-031640-I. J.L.-B. is funded by the Spanish Ministry of Science and Universities (MICIU/AEI/10.13039/501100011033) and NextGenerationEU/PRTR grants PID2019-107061GB-C61, PID2021-125627OB-C31, CNS2023-144309, and PID2023-150468NB-I00.
    This research made use of the following software: \texttt{astropy}, (a community-developed core Python package for Astronomy, \citealt{2022ApJ...935..167A}), \texttt{SciPy} \citep{2020NatMe..17..261V}, \texttt{matplotlib} (a Python library for publication quality graphics \citealt{2007CSE.....9...90H}), \texttt{numpy} \citep{2020Natur.585..357H} and \texttt{Tensorflow} \citep{tensorflow2015large}.
    This research has made use of data from the European Space Agency (ESA) mission
    {\it Gaia} (\url{https://www.cosmos.esa.int/gaia}), processed by the {\it Gaia}
    Data Processing and Analysis Consortium (DPAC,
    \url{https://www.cosmos.esa.int/web/gaia/dpac/consortium}). Funding for the DPAC
    has been provided by national institutions, in particular the institutions
    participating in the {\it Gaia} Multilateral Agreement. We also made use of NASA's Astrophysics Data System (ADS) Bibliographic Services, the SIMBAD database, operated at CDS, and the NASA Exoplanet Archive, which is operated by the California Institute of Technology, under contract with the National Aeronautics and Space Administration under the Exoplanet Exploration Program. We also used the data obtained from or tools provided by the portal exoplanet.eu of The Extrasolar Planets Encyclopaedia.
    \end{acknowledgements}

    \newpage
    \bibliographystyle{aa}
    \bibliography{aa55598-25}
    \newpage
    \begin{appendix}
    \nolinenumbers
    \section{Adding photometric and spectroscopic information}\label{appendix:additional_data}
    
    To add photometric and spectroscopic information for every simulated source we first built two differentiated samples, one corresponding to the single stars (negative case) and another for the binary stars (positive case). Each sample contained a combination of 50000 main sequence and giant real stars. In the absence of a large enough sample of bona fide single stars to construct our single stars subset, we resorted to the Gaia DR3 FGKM golden sample \citep{2023A&A...674A..39G}. These are set of about three million sources with very-high-quality photometry, astrometry, and stellar parameters. From the full FGKM sample we selected sources with no appreciable signs of binarity or multiplicity. To do this we required well-observed single-star solutions with a low \texttt{ruwe} ($\texttt{ruwe}\le1.2$ \citealt{2018A&A...616A...2L, 2021A&A...649A...5F}) and at least 11 visibility periods. We also added further filtering on \texttt{ipd\_frac\_multi\_peak} to avoid double stars with a large separation ($\gtrsim$1.5 arcsecond. \citealt{2021A&A...649A...5F}) and \texttt{ipd\_gof\_harmonic\_amplitude} to also reject pairs with smaller separations. Finally, to minimize contamination from binaries we excluded all sources listed in the Gaia DR3 Non-Single Star (NSS) catalogue \citep{ 2023A&A...674A...9H,2025A&A...693A.124G} (astrometric and spectroscopic binaries) using the \texttt{non\_single\_star} flag. We joined the resulting subset with the astrophysical parameters DR3 table \texttt{gaiadr3.astrophysical\_parameters} to obtain the mass estimate for each source. This can be translated into the following query in the Gaia archive:
    \newline
    \newline
    \texttt{SELECT TOP 100000 A.*, B.* FROM gaiadr3.gold\_sample\_fgkm\_stars AS B} 
    \\\texttt{JOIN gaiadr3.gaia\_source AS A USING (source\_id)}
    \\\texttt{WHERE A.ruwe <1.2}
    \\\texttt{AND A.visibility\_periods\_used > 11}
    \\\texttt{AND A.ipd\_frac\_multi\_peak < 2}
    \\\texttt{AND A.ipd\_gof\_harmonic\_amplitude < 0.1}
    \\\texttt{AND A.non\_single\_star = 0}
    \newline
    \newline
    \noindent For the binary stars we did the opposite operation, selecting astrometrically well-observed sources from the main Gaia DR3 table with high \texttt{ruwe} value. We removed the cuts on initial parameter determination (IPD) parameters and set the \texttt{non\_single\_star} flag to value 1, which indicates that the source is identified as an astrometric binary in the DR3 NSS catalogue. This can be translated into the following query in the Gaia archive:
    \newline
    \newline
    \texttt{SELECT TOP 100000 A.*, B.* FROM gaiadr3.gaia\_source AS A}
    \\\texttt{JOIN gaiadr3.astrophysical\_parameters AS B USING (source\_id)}
    \\\texttt{WHERE A.ruwe >1.4}
    \\\texttt{AND A.visibility\_periods\_used > 11}
    \\\texttt{AND A.non\_single\_star = 1}
    \newline
    \newline
    \noindent A description of all the parameters used in these queries is available in the online Gaia documentation\footnote{Gaia DR3 main source table description: \url{https://gea.esac.esa.int/archive/documentation/GDR3/Gaia_archive/chap_datamodel/sec_dm_main_source_catalogue/ssec_dm_gaia_source.html}}.
    \begin{figure}[!h]
    \centering
    \includegraphics[width=9cm]{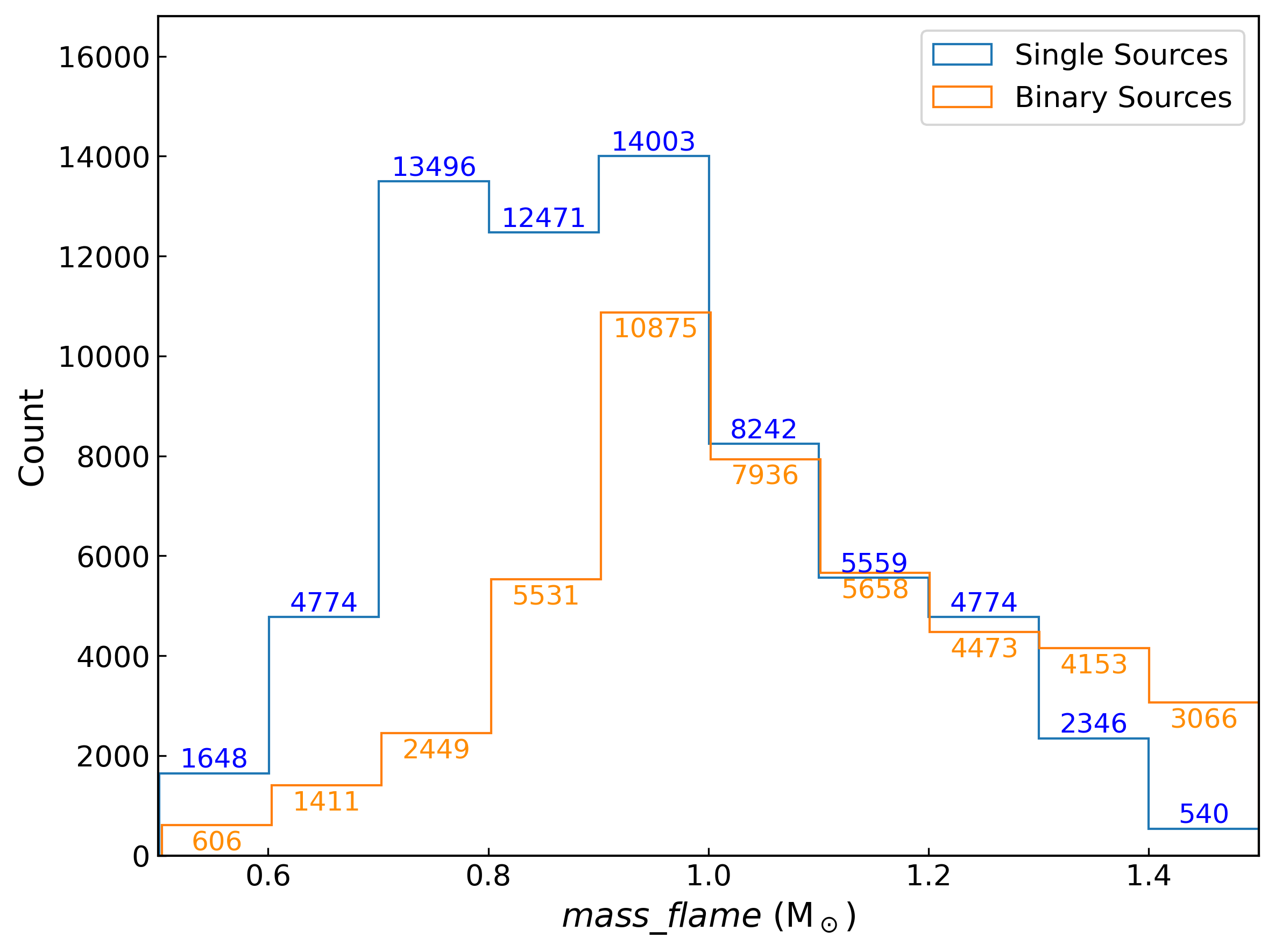}
    \vspace{10pt}
    \includegraphics[width=8.5cm]{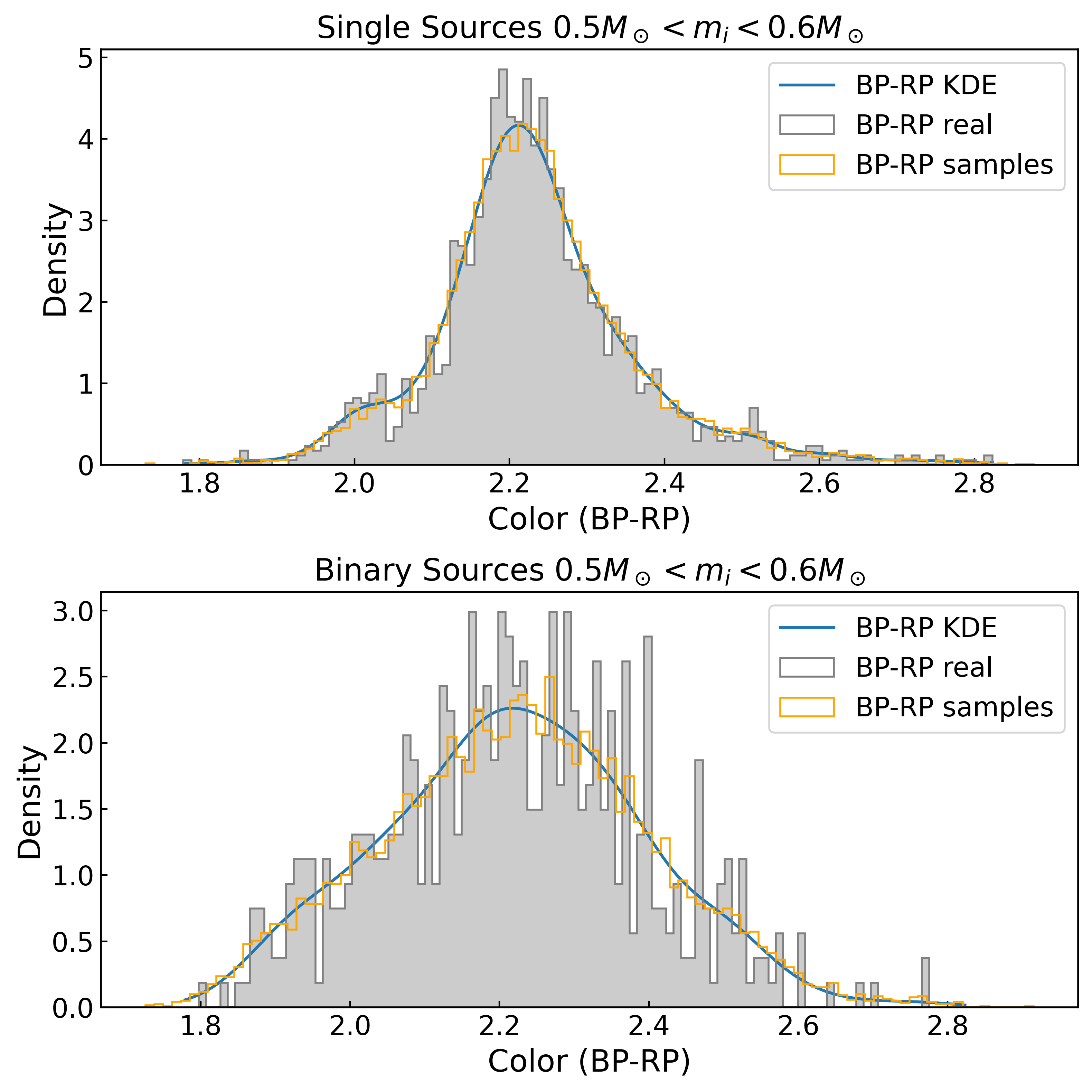}
    \caption{TOP: Mass distribution of single (blue) and binary (orange) samples with bins of 0.1$M_\odot$ on the variable \texttt{mass\_flame}. MIDDLE: Distribution of color indices for single sources (gray) compared to sampled colors (orange), obtained from a 1D KDE fit corresponding to the low end of the simulation mass range. BOTTOM: Equivalent plot as in the middle but for the binary sources in this case.}
    \label{fig:kde_fitting}
    \end{figure}
    For each generated sampled we only retained those stars that have a value of the \texttt{mass\_flame}\footnote{A description of this parameter can be found in the Gaia DR3 astrophysical parameters table: \url{https://gea.esac.esa.int/archive/documentation/GDR3/Gaia_archive/chap_datamodel/sec_dm_astrophysical_parameter_tables/ssec_dm_astrophysical_parameters.html}} parameter that falls within our simulated primary mass range of ($0.5 M_\odot\!\le\!M_{flame}\!\le\!1.5M_\odot$). The \texttt{mass\_flame} provides an estimate of the stellar mass and comes from the FLAME module of the Astrophysical Parameters Inference System  \citep[Apsis,][]{2023A&A...674A..26C}. For single stars this estimated mass refers to the mass of the star, and for binaries it corresponds to the mass of the primary object. Then, to generate synthetic stellar parameters for each star (either single or binary) we used a 2D kernel density estimation (KDE) approach \citep[e.g.,][]{Silverman:1070306,2015mdet.book.....S}. This allowed us to conditionally sample values from a 2D KDE fit of the joint probability density function $p(c,X)$ where $c$ is the Gaia $BP-RP$ color of a source and $X$ is a given DR3 parameter. Since our samples contain a combination of dwarf, sub-giants and giants, we considered that a non-parametric fitting approach such as KDE was more adequate to handle the potential multi-modality present in certain stellar parameters, and opted for kernel density estimation versus parametric distribution fitting. The average number of sources per $0.1M_\odot$ mass bin in the range ($0.6 M_\odot\!\le\!M_{flame}\!\le\!1.4M_\odot$) was in the order of thousands, see Figure~\ref{fig:kde_fitting} (top panel), so this guaranteed that we had sufficient statistics for the KDE sampling. The generation of synthetic parameters was done as follows:
    
    \begin{enumerate}
    \item We first divided each sample into mass bins of size 0.1$M_\odot$ using the parameter \texttt{mass\_flame}.

    \item Then, for each star in the given sample and mass bin we extracted the pairs ($c_i, X_i$), where $c_i$ is the \texttt{bp\_rp} color and $X_i$ is the value of the stellar parameter of interest (e.g., magnitude) for the $i$-th star in the sample.
    \item We then constructed a 2D kernel density estimator over the pairs $(c_i, X_i)$ to estimate the joint probability density function $p(c, X)$:
    \begin{equation}
    \hat{p}(c, X) = \frac{1}{N} \sum_{i=1}^{N} K\big((c - c_i, X - X_i); H\big),
    \end{equation}
    where $K(x; H)$ is a Gaussian bivariate kernel and $H$ is a so-called bandwidth matrix. The bandwidth matrix $H$ controls the smoothness of the KDE, and is defined as :
        \begin{equation}
        H = 
        \begin{bmatrix} 
        h_c^2 & h_{cX} \\ 
        h_{cX} & h_X^2 
        \end{bmatrix}, 
        \quad h_j \propto \sigma_j , \quad j \in \{c, X\},
        \end{equation}
        where $h_j$ is proportional to the standard deviation of the corresponding variable $\sigma_j$ and the off-diagonal elements $h_{cX}$ capture the correlation between color, $c$, and the variable, $X$. In practice $K(x; H)$ is automatically computed using \texttt{gaussian\_kde} from \texttt{scipy.stats}.
    
    \item Now, for any desired color, $c^\ast$, we compute the conditional probability distribution of $X$ as
    \begin{equation}
    \hat{p}(X \mid c^\ast) = \frac{\hat{p}(c^\ast, X)}{\int \hat{p}(c^\ast, X) \, dX}.
    \end{equation}
    which yields a smooth, continuous estimate of $p(X \vert c)$ for any value of $c$.
  
    \end{enumerate}
    \noindent This procedure generated a library of $M \times N$ joint probability density functions $p(X|c)$ where $M$ is the number of mass bins and $N$ is the number of synthetic parameters. Still, to use these joint PDFs, we needed to generate the missing value of the Gaia BP-RP color parameter for a given star with simulated mass $M_\star$. To do this we used the \texttt{isochrone} software package \citep{2015ascl.soft03010M} and performed a basic 1D interpolation using a main-sequence MIST isochrone \citep{2012MNRAS.427..127B}. We assumed solar metallicity and fixed 4.6GYr age for every star in our simulated samples, even though we now this is only roughly valid for a subset of the stars we used to generate the 2D KDEs library. However, given that: (1) the influence of photometry and spectroscopy parameters in the model decision is of second-order importance compared to the astrometric fit statistics, and (2) preserving correlations between parameters is of higher importance in this scenario than a high intrinsic parameter accuracy, we opted for this overall process of synthetic parameters generation versus one based on full usage of physical stellar models. Since the primary goal was to generate a plausible set of photometric and spectroscopic parameters, we considered that this approach was sufficient.

\begin{figure*}[!h]
    \centering
    \includegraphics[width=13.cm]{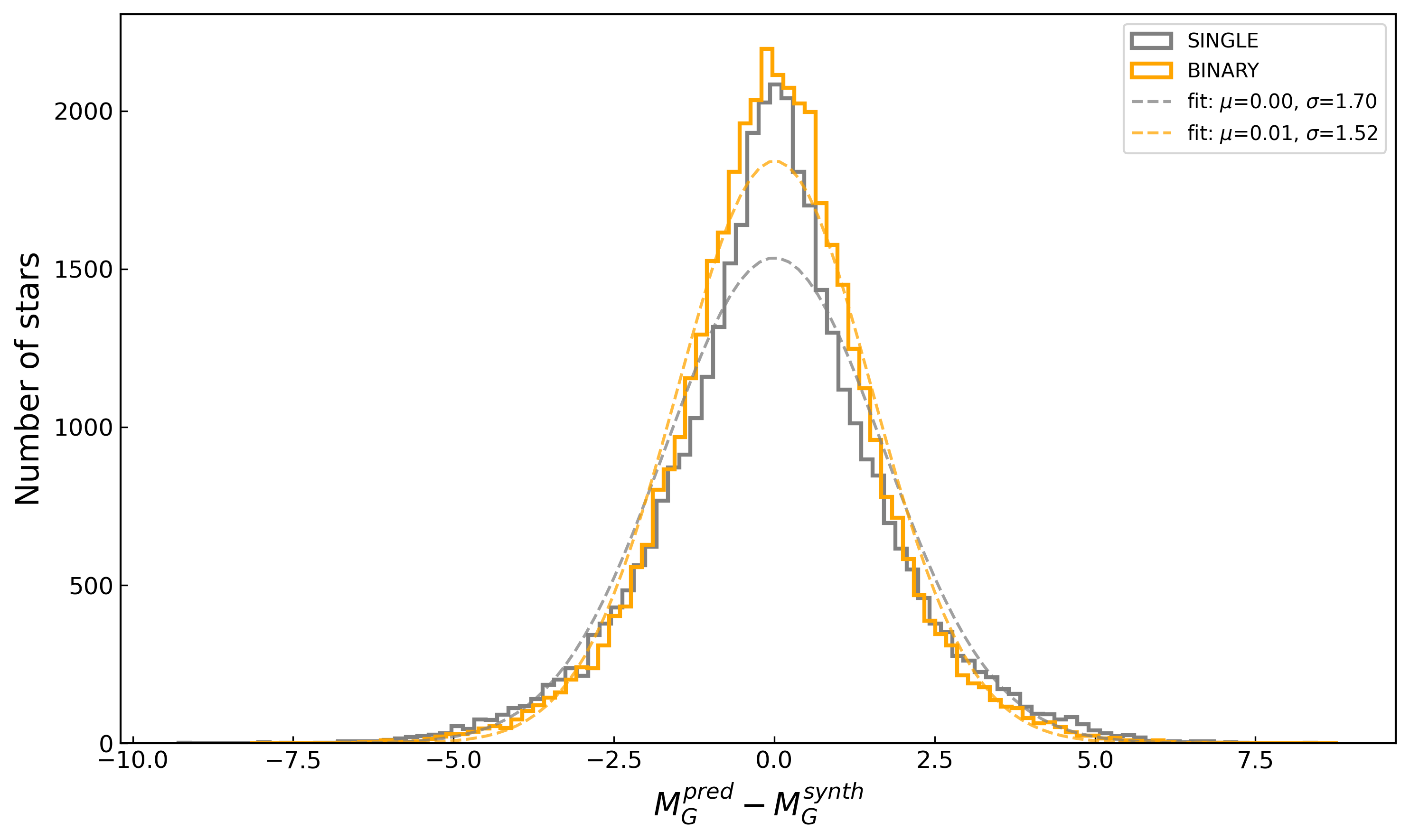}
    \vspace{20pt}
    \includegraphics[width=8cm]{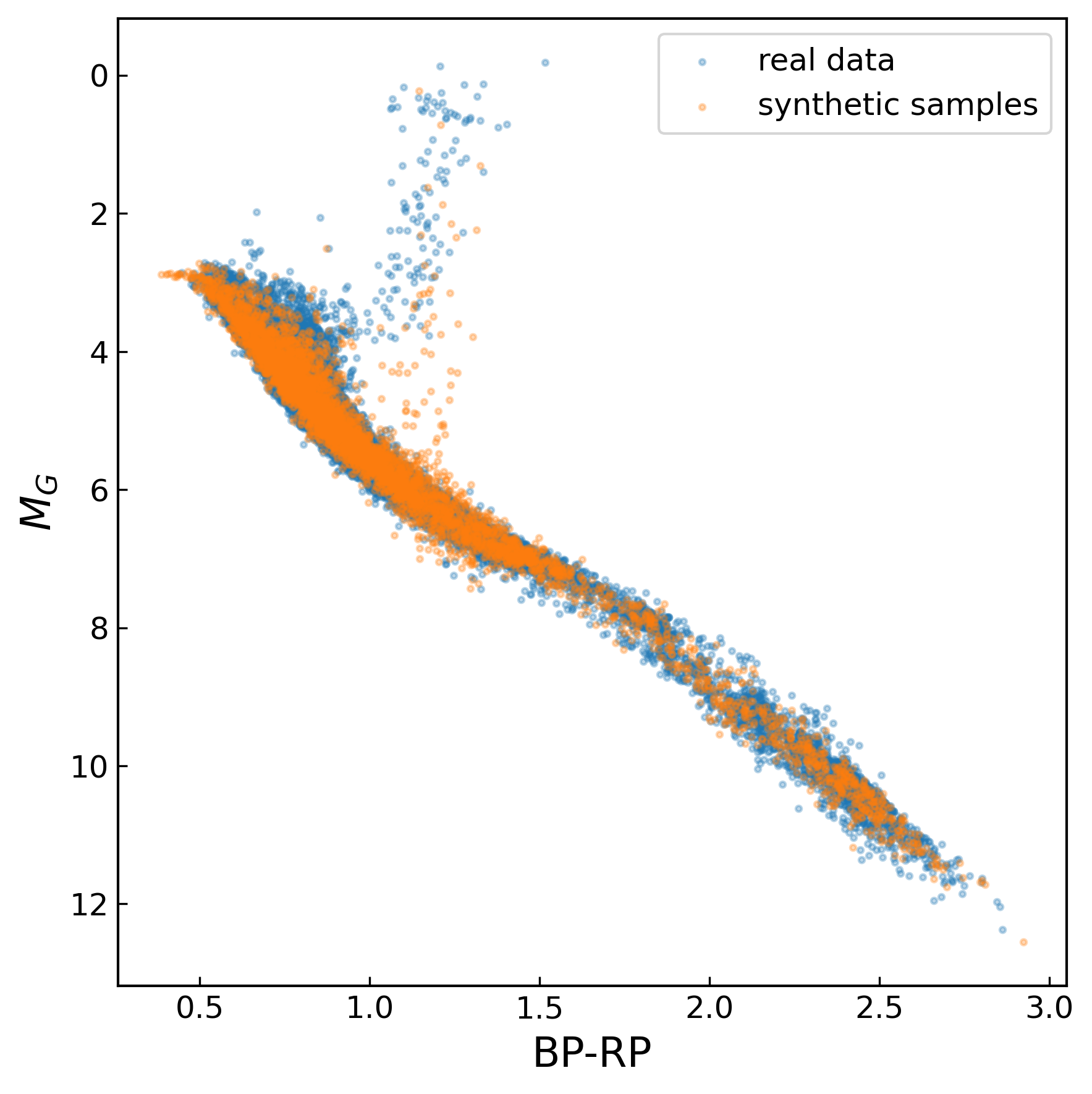}
    \hspace{2pt}
    \includegraphics[width=8cm]{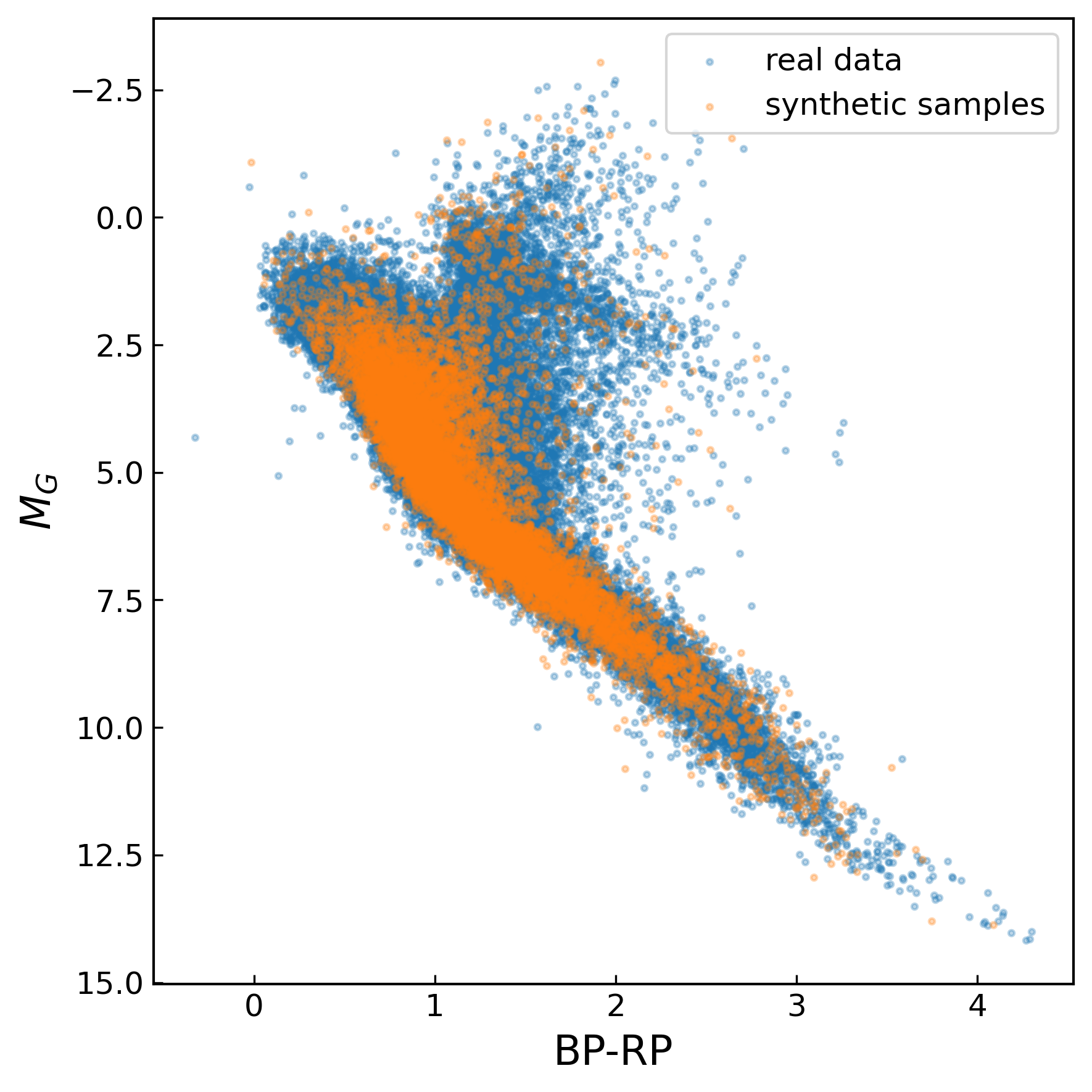}
    
    \caption{TOP: Distribution of the residuals $M^{pred}_G-M^{synth}_G$ for single and binary samples with corresponding Gaussian fits to the residuals. BOTTOM LEFT: Colour-magnitude diagram of the 100000 single (blue) sources and synthetic color-magnitude diagram with conditionally sampled color and magnitude values (orange). BOTTOM RIGHT: equivalent plot for the binaries with real binaries in (blue) and synthetic color-magnitude diagram with conditionally sampled color and magnitude values (orange).}
    \label{fig:example_params}
    \end{figure*}

    Nevertheless, to avoid any possible unphysical parameter combinations in our generated synthetic photometry, we implemented a coherence test that evaluates whether a simulated set of Gaia ($BP-RP$) color, absolute Gaia magnitude ($M_G$), and mass ($M_\star$) values are consistent with the physical properties of a single main-sequence star. To do this, we estimated the absolute $G$-band magnitude of a star with mass, $M_\star$, and color, $BP-RP$, using the following expressions,
    \begin{equation}
    M_G^{\mathrm{pred}}(M_\star, BP-RP) 
    = M_{\mathrm{bol}}(M_\star) - BC_G(BP-RP)
    ,\end{equation}
    where $M_{\mathrm{bol}}$ is the bolometric magnitude associated with a simulated stellar mass $M_\star$, and $BC_G$ is the bolometric correction in the Gaia $G$ band. The bolometric magnitude of the simulated star was computed as
    \begin{align}
        M_{\mathrm{bol}}(M_\star) 
        &= M_{\mathrm{bol},\odot} - 2.5 \log_{10}\!\left(\frac{L(M_\star)}{L_\odot}\right),
    \end{align}
    where $M_{\mathrm{bol},\odot} \simeq 4.74$ and $L(M_\star) \simeq M_\star^{4.0}$ \citep{2015AJ....149..131E}. The Gaia G-band bolometric correction ($BC_G$) was determined by interpolating the empirical calibration grid presented in \citep[][Table 6]{2020MNRAS.496.3887E}. In this work the authors provide empirically derived bolometric corrections as a function of intrinsic stellar parameters for solar-type main-sequence stars. We then checked the difference between the predicted absolute magnitude for a given star and the synthetically generated one and only retained those synthetic values for which this difference was lower than a threshold: $
    \lvert M_G^{\mathrm{pred}} - M_G^{\mathrm{synth}} \rvert < \tau =2 \sigma\;mag$, where $\sigma$ is the standard deviation of the computed residuals. We show in Figure~\ref{fig:example_params} the distribution of residuals $M_G^{pred} - M_G^{synth}$ and a comparison of the synthetic color-magnitude diagram obtained from our 2D KDE sampling procedure for both single and binary samples.
    
    Now, while our selection procedure for single stars is designed to produce a sample of astrometrically well behaved (this is single) stars, this sample could still be contaminated with short-period binaries, near-equal-mass binaries with face-on orbits, or long-period binaries with a longer period than the DR3 time baseline. To assess the possible impact of introducing wrongly labelled binaries as single systems, we consider the $f1$-score metric definition,  
    
    \begin{equation}
        f1=\frac{2 \cdot \text{Precision} \cdot \text{Recall}}{\text{Precision}+\text{Recall}}
        = \frac{\text{TP}}{\text{TP}+\tfrac{1}{2}(\text{FP}+\text{FN})},
        \label{eq:f1_score}
    \end{equation}
    
    \noindent where TP refers to true positives, FP to false positives, and FN to false negatives. Since mislabelling binaries as single stars effectively increases FN, the $f1$-score is expected to decrease as a consequence. We can analyse how a small perturbation $\Delta\text{FN}$ will affect the sensitivity of the $f1$-score. If we hold TP and FP fixed and take the partial derivative with respect to FN:
    \begin{equation}\label{eq:sensitivity}
        \frac{\partial f1}{\partial \text{FN}}
        = -\,\frac{1}{2}\,\frac{\text{TP}}{\bigl(\text{TP}+\tfrac{1}{2}(\text{FP}+\text{FN})\bigr)^2}.
    \end{equation}
    Hence, for a small change $\Delta\text{FN}$,
    \begin{equation} \label{eq1}
    \begin{split}
        \Delta f1 \approx -\,\frac{1}{2}\,\frac{\text{TP}}{\bigl(\text{TP}+\tfrac{1}{2}(\text{FP}+\text{FN})\bigr)^2}\,\Delta\text{FN},
        \\
        \frac{\Delta f1}{f1}\approx -\,\frac{\Delta\text{FN}}{2\bigl(\text{TP}+\tfrac{1}{2}(\text{FP}+\text{FN})\bigr)}.
    \end{split}
    \end{equation}
    The magnitude of the drop in $f1$-score depends mainly on the class balance. Using the derived formulae, we can analyse the expected degradation of performance for a balanced, strong model such as our ExoDNN, when a small fraction of FN is added. For example, assuming $\text{TP}=900,\ \text{FP}=100,\ \text{FN}=100$, and using equation \ref{eq:f1_score} we compute the $f1$-score as: 
    \[
    f1=\frac{900}{900+\frac{1}{2}(100+100)}=0.9.
    \]
    The sensitivity in this score is then given by Equation \ref{eq:sensitivity}:
    \[
    \frac{\partial f1}{\partial \text{FN}}
    = -\frac{1}{2}\,\frac{900}{1000^2}
    =-0.00045.
    \]
    Therefore, if we increase $\Delta \text{FN}=10$ (i.e.,\ $+10\%$), using Equation \ref{eq1} we have:
    \[
    \Delta f1 \approx -0.00045\times 10 = -0.0045,
    \qquad
    f1^\prime \approx 0.8955,
    \]
    and we expect an absolute drop of $0.0045$ in the $f1$-score in this case.  For $\Delta \text{FN}=5$, the drop is halved, $f1^\prime\approx 0.8978$. We can see how for a balanced model the impact of introducing a small fraction ($\leq$10\%) of false negatives is relatively small.
    
    \section{Additional models}\label{appendix:additional_models}

    For a relative performance comparison against ExoDNN, the selected binary classifiers were logistic regression and random forest. The logistic regression is similar to our network in the sense that is a probabilistic classifier whose objective is to minimize the log-loss, but it is fundamentally a linear model which cannot capture complex non-linear relationships unlike our ExoDNN. Random Forest, on the other side, is a non-parametric approach that uses an ensemble of decision trees to learn decision boundaries via recursive feature splits. We trained both classifiers using the same training data used to train ExoDNN and applied equivalent pre-processing steps, namely, robust scaling of the data before training. Table~\ref{tab:logistic_regression} and Table~\ref{tab:random_forest} list the different hyper-parameters used to train the alternative classifiers.
    
    \begin{table}[ht]
    \caption{Best hyper-parameters for logistic regression}
    \centering
    \label{tab:logistic_regression}
    \begin{tabular}{|l|c|}
    \hline
    \textbf{Hyper-parameter} & \textbf{Value} \\
    \hline
    C & 0.8 \\
    max\_iter & 100 \\
    solver & newton-cg \\
    tol & 0.0001 \\
    \hline
    \end{tabular}
    \end{table}
    
    \begin{table}[ht]
    \caption{Best hyper-parameters for random forest}
    \label{tab:random_forest}
    \centering
    \begin{tabular}{|l|c|}
    \hline
    \textbf{Hyper-parameter} & \textbf{Value} \\
    \hline
    bootstrap & true \\
    max\_depth & 1 \\
    max\_features & 4 \\
    min\_samples\_leaf & 1 \\
    min\_samples\_split & 2 \\
    n\_estimators & 50 \\
    \hline
    \end{tabular}
    \end{table}
     
    \section{Model input selection}\label{appendix:model_input_data}
    
\begin{figure}[!h]
    \centering
    \includegraphics[width=9cm]{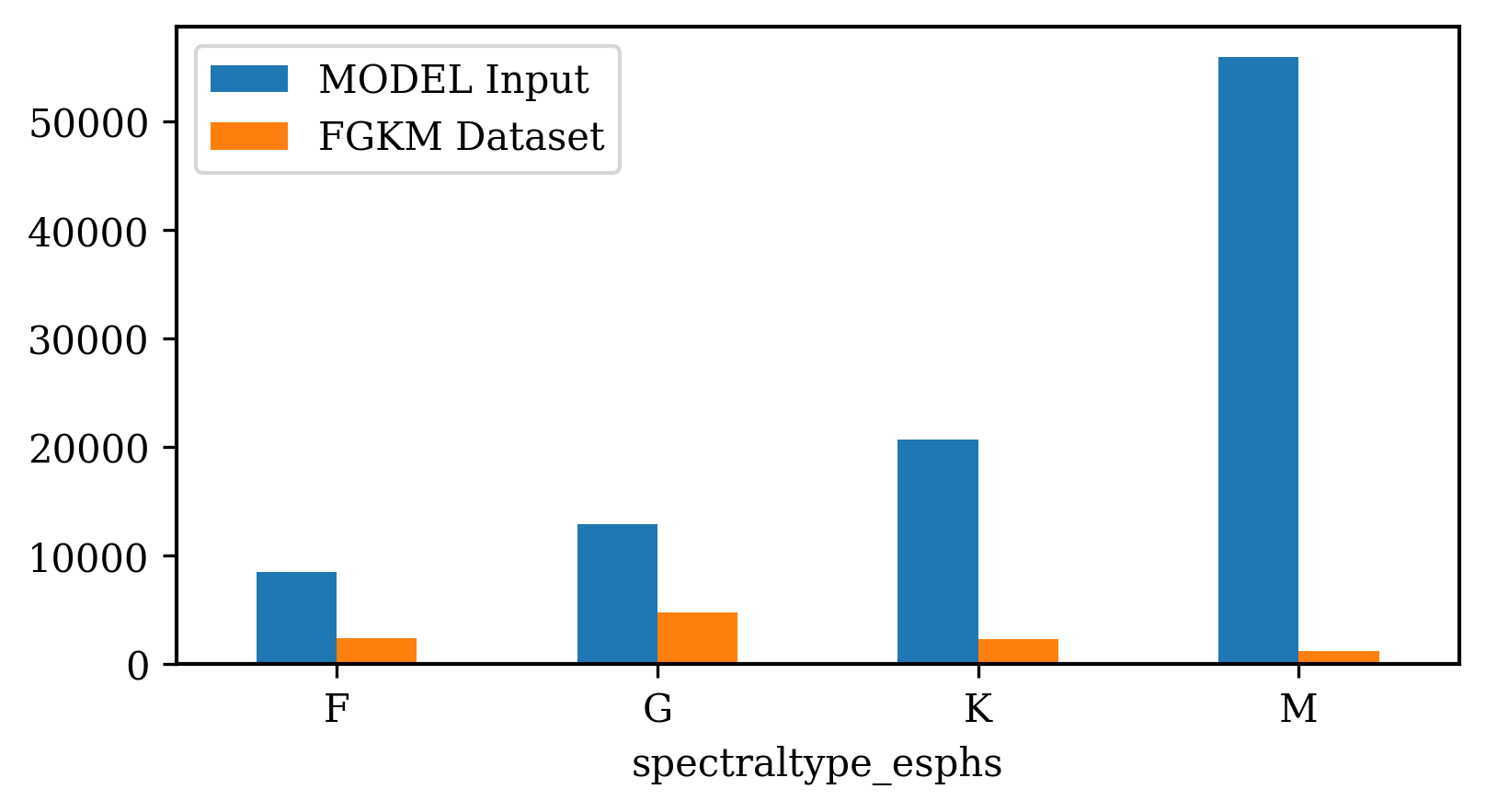}
    \caption{Comparison of the number of sources per spectral type between the inputs used to make model prediction and the FGKM golden sample \citep{2023A&A...674A..26C}}
    \label{fig:model_inputs}
    \end{figure}

    To generate the input for the deep learning model prediction we selected sources from the Gaia main source table (\texttt{gaiadr3.gaia\_source}) in the Gaia archive via the following query:
    \newline
    \texttt{SELECT A.*, B.* FROM gaiadr3.gaia\_source AS A} 
    \\\texttt{JOIN gaiadr3.astrophysical\_parameters AS B USING (source\_id)}
    \\\texttt{WHERE A.parallax >10}
    \\\texttt{AND A.visibility\_periods\_used>11}
    \\\texttt{AND A.teff\_gspphot > 2500}
    \\\texttt{AND A.teff\_gspphot < 7500}
    \\\texttt{AND A.phot\_variable\_flag NOT LIKE 'VARIABLE'}
    \newline
    \noindent The join with the astrophysical parameters DR3 table (\texttt{gaiadr3.astrophysical\_parameters}) is required to obtain estimated stellar parameters for each source generated by the astrophysical parameters inference system  \citep[Apsis,][]{2023A&A...674A..26C}. Figure~\ref{fig:model_inputs} shows the number of sources retrieved by the query with a breakdown per spectral type.
    
    \end{appendix}
    
\end{document}